\documentstyle[aps,epsf]{revtex}   

\begin{document}

\title{Comparison of various models of particle multiplicity distributions
using a general form of the grand canonical partition function}

\author{S.J. Lee}
\address{Department of Physics and Institute of Natural Sciences,
                    Kyung Hee University, Suwon, KyungGiDo, Korea}

\author{A.Z. Mekjian}
\address{Department of Physics, Rutgers University,
                    Piscataway, New Jersey}

\maketitle

\begin{abstract}

Various phenomenological models of particle multiplicity distributions
are discussed using a general form of the grand canonical partition 
function.
These phenomenological models include a wide range of varied processes
such as coherent emission or Poisson processes,
chaotic emission resulting in a negative binomial distribution,
combinations of coherent and chaotic processes called signal/noise
distributions, and models based on field emission from Lorentzian
line shapes leading to Lorentz/Catalan distributions.
These specific cases can be written as special cases of a more 
general distribution.
Using this grand canonical approach moments and cumulants, combinants, 
hierarchical structure, void scaling relations, KNO scaling features, 
clan variables and branching laws associated with stochastic or
ancestral variables are discussed.
It is shown that just looking at the mean and fluctuation of data is not
enough to distinguish these distributions or the underlying mechanism.
A generalization of the Poisson transform of a distribution
and the Poissonian decomposition of it into a compound 
or sequential process is also given.

\end{abstract}

\pacs{
PACS No.: 25.75.Dw, 25.75.Gz, 24.10.Pa, 05.30.Jp, 05.40.-a
 }

\section{Introduction}

Pion multiplicity distribution and their associated fluctuation and 
correlations have been studied at the CERN SPS \cite{ref1,ref2} in the past,
are being investigated at RHIC presently,
and in the future will be studied at the LHC.
Because of the very large number of pions
and other produced particles in very high energy collisions,
event-by-event studies can be carried out and
are important tools for many reasons.
One current reason for studying them resides in the hope that
anomalous fluctuations will remain from a transition of a
quark-gluon phase to hadronic phase,
thus offering one of the signals of the formation of the QG phase.
Several models predict large fluctuations such as the disoriented
chiral condensate \cite{ref3,ref4}
and in density fluctuations from droplets arising in a first order
phase transition \cite{ref5}.
A well known procedure for studying correlations uses the
Bose-Einstein symmetries associated with pions in
a Hanbury Brown-Twiss analysis \cite{ref6}.
Such an analysis gives information about the space time
history of the collision through measurements of source parameters.
If the density of pions becomes large, Bose-Einstein correlation
may also lead to a strongly emitting system which has been
called a pion laser \cite{ref7}.
The pion laser model has been recently solved analytically
by T. Cs\"{o}rg\"{o} and J. Zimanyi \cite{ref7a}.
The importance of Bose-Einstein correlations has also been
illustrated in the observation of a condensation of atoms in
a harmonic oscillator or laser trap \cite{ref8}.
Previous interest in pionic distributions have centered around
the possibility of intermittency behavior \cite{ref9} and
fractal structure based on parallels with turbulent flow in fluids.
A distribution widely used to discuss such features has been
the negative binomial (NB) distribution \cite{ref10}
with its associated clan structure \cite{ref11,ref12}
and KNO scaling feature \cite{ref13}.
KNO scaling properties have been interpreted in terms of a
phase transition associated with a Feynman-Wilson gas \cite{ref14}.
Various other issues associated with pions include evidence for 
thermalization \cite{ref15}, critical point fluctuations \cite{ref16,ref17},
fluctuations from a first order phase transition \cite{ref18},
charge particle ratios and question of chemical equilibrium \cite{ref19},
the behavior of fluctuations in net charge in a QG plasma for
transition \cite{ref20,ref21}.

For lower energy heavy ion collisions, multifragmentation of nuclei
takes place. The fragment distribution can also be described
statistically by considering all the possible partition of $A$
nucleons into smaller clusters \cite{frag,canon}. 
This study gives a tool
for the description of nuclear multifragmentation distributions \cite{massd}, 
nuclear liquid-gas phase transition \cite{lgpha}, 
critical exponent, intermittency, and chaotic behavior \cite{scale,power}
of nuclear multifragmentation.
The same model can describe pionic distribution .
This possibility arises in our approach which is based on Feynman path 
integral methods where symmetrization of bosons or anti-symmetrization
of fermions leads to a cycle class decomposition of the
permutations associated with these symmetries.
The correspondance comes from the identification of clusters of size $k$
and the cycles of length $k$ in a permutation as discussed below.

In next section a summary of the generalized statistical model
will be given. 
Various models of particle multiplicity distribution will
then be developed.
Moreover, we derive a generalized model
which can further be reduced to a geometric, negative binomial,
and Lorentz/Catalan model.
These various models used in describing pion and particle multiplicity
distributions then are examined with the generalized model in Sect. III.
In Sect. IV, we further compare various statistical properties between
different models within the generalized model.

\section{Generalized probability distribution}

\begin{figure}[hbt]

\begin{center}

\setlength{\unitlength}{0.8cm}  
\begin{picture}(25,8)(7.5,0.0)     

\thicklines

\newsavebox{\diagbox}
\savebox{\diagbox}(0.5,0.5){ }

\put(12.2,6.2){\large\bf a)}

\put(8.0,0.25){\large\bf $\vec n = (1^4,4^2,6^1,7^2,9^2)$}

\put(8.01, 1.51){\line(0,1){4.5}}
\multiput(8.0, 5.5)(0.5,0){2}{\frame{\usebox{\diagbox}}}
\multiput(8.0, 5.0)(0.5,0){2}{\frame{\usebox{\diagbox}}}
\multiput(8.0, 4.5)(0.5,0){4}{\frame{\usebox{\diagbox}}}
\multiput(8.0, 4.0)(0.5,0){5}{\frame{\usebox{\diagbox}}}
\multiput(8.0, 3.5)(0.5,0){5}{\frame{\usebox{\diagbox}}}
\multiput(8.0, 3.0)(0.5,0){7}{\frame{\usebox{\diagbox}}}
\multiput(8.0, 2.5)(0.5,0){7}{\frame{\usebox{\diagbox}}}
\multiput(8.0, 2.0)(0.5,0){7}{\frame{\usebox{\diagbox}}}
\multiput(8.0, 1.5)(0.5,0){11}{\frame{\usebox{\diagbox}}}
\put(8.99, 5.0){\line(0,1){1.0}}
\put(9.99, 4.5){\line(0,1){0.5}}
\put(10.49, 3.5){\line(0,1){1.0}}
\put(11.49, 2.0){\line(0,1){1.5}}
\put(13.49, 1.5){\line(0,1){0.5}}
\put(8.01, 1.51){\line(1,0){5.5}}
\put(11.51, 2.01){\line(1,0){2.0}}
\put(10.51, 3.51){\line(1,0){1.0}}
\put(10.01, 4.51){\line(1,0){0.5}}
\put(9.01, 5.01){\line(1,0){1.0}}
\put(8.01, 6.01){\line(1,0){1.0}}
\multiput(7.5,6.5)(0,0.01){2}{\vector(1,0){0.5}}
\multiput(9.5,6.5)(0,0.01){2}{\vector(-1,0){0.5}}
\put(8.0,6.2){\makebox(1.0,0.6){\large\bf $n_k$}}
\multiput(7.6,3.2)(0,0.01){2}{\vector(0,-1){1.7}}
\multiput(7.6,4.3)(0,0.01){2}{\vector(0,1){1.7}}
\put(7.4,3.5){\makebox(0.3,0.6)[r]{\large\bf $k$}}
\put(15.5,0){
\setlength{\unitlength}{0.8cm}  
\begin{picture}(10,8)(0,0)
\thicklines

\savebox{\diagbox}(0.5,0.5){ }

\put(8,6.2){\large\bf b)}

\put(3.5, 7.0){\frame{\usebox{\diagbox}}}
\put(4.2,7.3){$1$}
\put(3.6,6.9){\vector(-3,-2){1.05}}
\put(3.9,6.9){\vector(2,-3){0.85}}

\multiput(2.0, 5)(0,0.5){2}{\frame{\usebox{\diagbox}}}
\put(2.7,5.6){$2$}
\put(2.05,4.9){\vector(-2,-1){0.65}}
\put(2.4,4.9){\vector(3,-2){1.1}}

\multiput(4.5, 5.0)(0.5,0){2}{\frame{\usebox{\diagbox}}}
\put(5.7,5.4){$1^2$}
\put(4.52,4.9){\vector(-1,-1){0.78}}
\put(5.45,4.9){\vector(3,-4){0.95}}

\multiput(1., 3.0)(0,0.5){3}{\frame{\usebox{\diagbox}}}
\put(1.7,4.0){$3$}
\put(1.1,2.9){\vector(-1,-1){0.75}}
\put(1.4,2.9){\vector(1,-4){0.3}}

\multiput(3.5, 3.0)(0,0.5){2}{\frame{\usebox{\diagbox}}}
\put(4.0, 3.0){\frame{\usebox{\diagbox}}}
\put(4.2,3.8){$(1,2)$}
\put(3.6,2.9){\vector(-4,-3){1.6}}
\put(3.9,2.9){\vector(0,-1){1.75}}
\put(4.4,2.9){\vector(3,-4){1.3}}

\multiput(6.0, 3.0)(0.5,0){3}{\frame{\usebox{\diagbox}}}
\put(7.1,3.7){$1^3$}
\put(6.4,2.9){\vector(-1,-3){0.6}}
\put(7.3,2.9){\vector(1,-2){1.1}}

\multiput(0., 0.0)(0,0.5){4}{\frame{\usebox{\diagbox}}}
\put(0.7,1.6){$4$}

\multiput(1.5, 0.0)(0,0.5){3}{\frame{\usebox{\diagbox}}}
\put(2.0, 0.0){\frame{\usebox{\diagbox}}}
\put(2.15,1.1){$(1,3)$}

\multiput(3.5, 0.0)(0,0.5){2}{\frame{\usebox{\diagbox}}}
\multiput(4.0, 0.0)(0,0.5){2}{\frame{\usebox{\diagbox}}}
\put(4.7,0.7){$2^2$}

\multiput(5.5, 0.0)(0.5,0){3}{\frame{\usebox{\diagbox}}}
\put(5.5, 0.5){\frame{\usebox{\diagbox}}}
\put(6.2,0.8){$(1^2,2)$}

\multiput(8.0, 0.0)(0.5,0){4}{\frame{\usebox{\diagbox}}}
\put(9.6,0.7){$1^4$}

\end{picture}
}

\end{picture}

\caption{Building partitions with blocks.
}
  \label{fig1}

\end{center}

\end{figure}
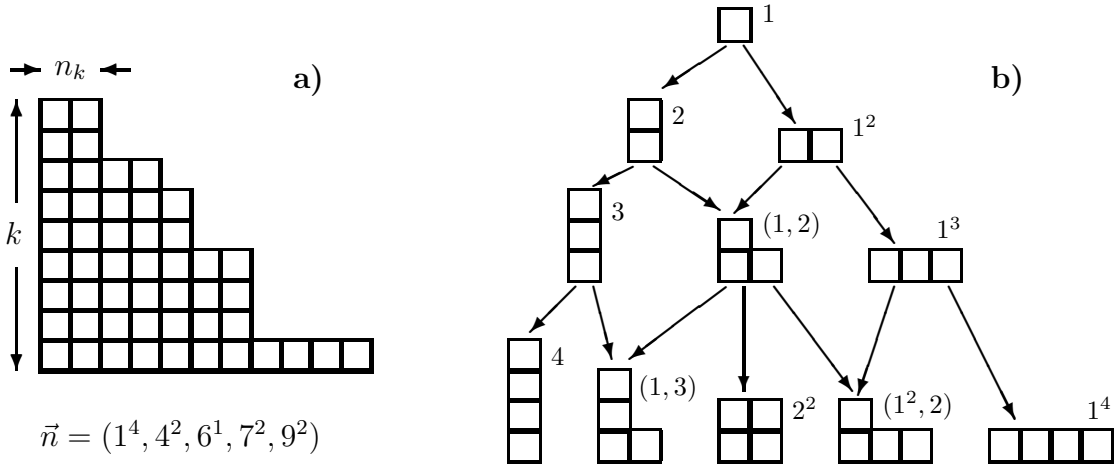

Consider a system composed of $N$ different types of species
or objects which could be the fragments 
in a fragmentation or in a cycle class description.
Any event of such a system can be associated with a vector
 $\vec n = \{n_k\} = (n_1, n_2, \cdots, n_k, \cdots, n_N)$
or $1^{n_1} 2^{n_2} 3^{n_3} \cdots k^{n_k} \cdots N^{n_N}$
where the non-negative integer $n_k$ is the number of individuals 
of species $k$.
For example $n_k$ can be the number of clusters of size $k$
or the number of cycles of length $k$ in a given permutation
of $n$ particles.
The later is important for Bose-Einstein and Fermi-Dirac statistics.
A general block picture of $\vec n$ is shown in Fig.\ref{fig1}a.
Fig.\ref{fig1}b shows how the various partition can be developed
as an evolution from successively smaller systems.
The multiplicity, i.e., the total number of individuals is then
\begin{eqnarray}
 M &=& \sum_{k=1}^N n_k    \label{multip} 
\end{eqnarray}
The number of species $N$ can be infinity in general.

Various probability distributions related with this system
can be developed by assigning an appropriate weight $x_k$ to
each individual of type $k$ species.
A weight $W(\vec x, \vec n)$ is then given to each event $\vec n$
and the type of weight that will be considered has the structure;
\begin{eqnarray}
 W(\vec x, \vec n) &=& \prod_{k=1}^N \left[\frac{x_k^{n_k}}{n_k!}\right]
      \label{weight}
\end{eqnarray}
The $n_k!$ are Gibbs factors. The $x_k$ will be given below and
contains various physical quantities.
Summing the weight $W(\vec x, \vec n)$ over all the possible
events of $\vec n$, the grand canonical partition function $Z(\vec x)$
of the system is given as
\begin{eqnarray}
 Z(\vec x) &=& \sum_{\vec n} W(\vec x, \vec n)
   = \sum_{\vec n} \prod_{k=1}^N \left[\frac{x_k^{n_k}}{n_k!}\right]
   = \exp\left[\sum_{k=1}^N x_k\right]
        \label{grandptf}  
\end{eqnarray}
The last equation holds due to the form of Eq.(\ref{weight})
of the weight $W(\vec x, \vec n)$, i.e., the factor $x_k^{n_k}/n_k!$ 
is the $n_k$-th order expansion term of $e^{x_k}$ .

Introducing other quantities $\alpha_k$ to each individual entity
or group of type $k$,
 $\vec\alpha = \{\alpha_k\}
     = (\alpha_1, \alpha_2, \cdots, \alpha_k, \cdots, \alpha_N)$,
we can define a canonical partition function $Z_A(\vec x)$ with a fixed $A$ as
\begin{eqnarray}
 A &=& \sum_{k=1}^N \alpha_k n_k = \vec\alpha\cdot\vec n
            \label{asum}  \\
 Z_A(\vec x) &=& \sum_{\vec n_A} W(\vec x, \vec n)
   = \sum_{\vec n_A} \prod_{k=1}^N \left[\frac{x_k^{n_k}}{n_k!}\right]
            \label{canoptf}   
\end{eqnarray}
with $Z(\vec x) = \sum_A Z_A(\vec x)$.
Here $\sum_{{\vec n}_A}$ is the summation over all events with
a fixed value of $A$, i.e., over a canonical ensemble of a fixed $A$,
and the $\sum_A$ is a summation over all the possible values of $A$;
it becomes $\sum_{A=0}^\infty$ for the case of $\alpha_k = k$ with
positive integer $k$.
The case $\alpha_k = k$ is encountered in fragmentation problems 
and permutation problems.
If we take $x_k \propto z^{\alpha_k}$, then the canonical partition
function $Z_A(\vec x)$ is the $z^A$ dependent term  
of the grand canonical partition function $Z(\vec x)$.
The physics of the canonical ensemble depends on the choice of
the quantity $\alpha_k$ \cite{frag,canon}.
There always is at least one event having $A = 0$, i.e., the
event where all $n_k$'s are zero, $\vec n = \vec 0$.
Thus $Z_0(\vec x) = 1$ if all $\alpha_k$ are non zero positive
since then there are no other possible events having $A = 0$.
Due to the form of the weight $W(\vec x, \vec n)$ given by Eq.(\ref{weight})
the canonical partition function $Z_A(\vec x)$  
satisfies a recurrence relation \cite{scale,zarecur}   
\begin{eqnarray}
 Z_A(\vec x) &=& \frac{1}{A} \sum_k \alpha_k x_k Z_{A-\alpha_k}(\vec x)
          \label{recur}
\end{eqnarray}
This relation is nothing but the constraint Eq.(\ref{asum})
in terms of the mean $<n_k>_A$ using Eq.(\ref{factma}).
For non-zero positive $\alpha_k$, there is no case having $A < 0$, i.e.,
$Z_A = 0$ for $A < 0$ 
and thus the $Z_A$ can be obtained by the recurence relation of
Eq.(\ref{recur}) starting from $Z_0(\vec x) = 1$.

To study the statistical properties of a system where the number of 
individuals is
independent of their species type, we can choose $\alpha_k = 1$.
For fragmentation, all the clusters are treated the same independent of
their size $k$ or internal structure by choosing $\alpha_k = 1$.
In pion distribution, $\pi^+$, $\pi^0$, and $\pi^-$ are treated
the same if we choose $\alpha_k = 1$ independent of their charge.
By choosing $\alpha_k = 1$, we can study jet distributions
without considering any further process of hadronization.
Then $A = \sum_k n_k = M$ and the canonical ensemble is a set 
of systems having the same number of individuals $A = M$ independent 
of their internal structure and the canonical partition function
$Z_A(\vec x) = Z_M(\vec x)$ is the expansion of grand canonical partition 
function $Z(\vec x)$ in terms of $z$ with a power of multiplicity $M$
and $z^A = z^M$ counts the total multiplicity of individuals in the system.
As we will see later, this uniform treatment of all the species
independent to their internal structure leads to a Poisson distribution.

If we take $\alpha_k = k$ with the $k$ to be a positive integer
representing the number of constituent particles 
(such as nucleons in a nuclear fragment 
or cycle length of a permutation),
then $A = \sum_k k n_k$ is the total number of the constituent
particles in the system and $z^A$ term in $Z(\vec x)$
is the canonical partition function $Z_A$ with the total of $A$ constituent
particles. For this case the weight variable $z$ is the variable
counting the number of constituent particles, i.e., the fugacity
of the constituent with $z = e^\mu$ and with $\mu$ the chemical
potential of the constituent particle.
This case is related to the nuclear multifragmentation model 
in Refs. \cite{frag,canon}
and various models of pion distribution discussed below.
For $\alpha_k = k$,
and if we also take $x_k = x$ then all the clusters are treated the same
independent of the size $\alpha_k = k$.
The weight $W \propto x^M$
is then the same for all partitions having the same multiplicity except
for the Gibb's factor.
Thus $x$ counts the multiplicity 
with $x$ being a fugacity of a cluster.
If we choose $x_k = z^{k}$ then the weight $W(\vec x, \vec n)$
is the same for all partitions having same $A$.
All the nucleons are treated the same independent of 
the cluster it belongs to thus counting the number of constituent 
nucleons with $z$ being a fugacity of the constituent.
If we choose $x_k = z^{\alpha_k}$ for a general value of $\alpha_k$
then the weight $W(\vec x, \vec n)$
is the same for all partitions having same $A$.
All the constituents 
per unit $\alpha_k$ are treated the same independent of 
the cluster it belongs to thus counting $A$ as the corresponding 
total quantity of $\alpha_k$ with $z$ being a 
fugacity of unit $\alpha_k$.

On the other hand, if we choose $\alpha_k$ to be the energy $\epsilon_k$
of species $k$ or $k$-th level, then
 $A = \sum_k \alpha_k n_k = \sum_k \epsilon_k n_k = E$
is the total energy of the system
and $Z_A(\vec x)$ becomes the partition function of a
canonical ensemble with a fixed total energy $E$.
If we choose $\alpha_k$ to be the charge $q_k$
of species $k$, then $A = \sum_k q_k n_k = Q$
is the total charge of the system.
Most discussions in this section applies to a general
value of $\alpha_k$. 
However we will concentrate more on the choice of $\alpha_k = k$
here for simplicity in notation.

\subsection{Probability distribution}

In a canonical ensemble of fixed $A$, we can define a probability
distribution of a specific partition $\vec n$ as
\begin{eqnarray}
 P_A(\vec x, \vec n) &=& \frac{W(\vec x, \vec n)}{Z_A(\vec x)}
\end{eqnarray}
With this probability, various mean values, fluctuations
and correlations of the number of species $n_k$ can be evaluated 
as a ratio of canonical partition functions for two different values 
of $A$ such as \cite{frag,canon}
\begin{eqnarray}
 \left<\frac{n_k!}{(n_k-m)!} \frac{n_j!}{(n_j-l)!}\right>_A
   &=& \sum_{\vec n_A} \frac{n_k!}{(n_k-m)!}
            \frac{n_j!}{(n_j-l)!} P_A(\vec x, \vec n)
    = x_k^m x_j^l \frac{Z_{A - m \alpha_k - l \alpha_j}(\vec x)}{Z_A(\vec x)}
          \label{factma}
\end{eqnarray}
Thus we have
 $<n_k>_A = x_k \frac{Z_{A-\alpha_k}(\vec x)}{Z_A(\vec x)}$, which shows
that the mean number of species $k$ in a canonical ensemble with
fixed $A$ is proportional to the weight $x_k$ assigned to the species
and the ratio of canonical partition function with different $A$.
The recurrence relation of Eq.(\ref{recur}) then follows simply
using the fact that $A = \sum_{k=1}^N \alpha_k <n_k>$. 
This distribution has been used in describing various fragment 
distributions in nuclear multifragmentation \cite{frag,canon,massd}.

Now knowing the partition functions $Z(\vec x)$ and $Z_A(\vec x)$,
we can associate a probability $P_A(\vec x)$ of the system to have a 
fixed value of $A$ in a grand canonical ensemble as
\begin{eqnarray}
 P_A(\vec x) &=& \frac{Z_A(\vec x)}{Z(\vec x)}
    = \frac{1}{Z(\vec x)} \frac{1}{\Gamma(A+1)}
      \left[\left(\frac{d}{d z}\right)^A Z(\vec x, z)\right]_{z=0} 
                 \label{pax}  \\
 Z(\vec x, z) &=& \sum_A Z_A(\vec x) z^A
    = \exp\left[\sum_k x_k z^{\alpha_k}\right]   \label{grandzxz}
\end{eqnarray}
The last step follows from the fact that 
the $z^A$ power term of $Z(\vec x, z)$ is $Z_A(\vec x)$
if we put $x_k = z^{\alpha_k}$.  
Thus the generating function $Z(\vec x, z)$ of $P_A$ can also
be looked at as a grand canonical partition function with
the weight $x_k$ replaced to be $x_k z^{\alpha_k}$,
where the variable $z$ counts $A$ explicitly
and $Z(\vec x, 1) = Z(\vec x)$.
If we consider $P_A(\vec x, z) = P_A(\vec x) z^A$ then $z$ has two roles;
one as a weight which is assigned the same to each constituent and 
another as a generating  parameter of the probability $P_A(\vec x)$.
For the case that $Z_0(\vec x) = 1$, the void probability $P_0$ \cite{hiera}
is the inverse of the grand canonical partition function, i.e.,
\begin{eqnarray}
 P_0(\vec x) &=& \frac{Z_0(\vec x)}{Z(\vec x)} = Z^{-1}(\vec x)
\end{eqnarray}
Inversely the canonical partition function $Z_A(\vec x)$ is the 
probability $P_A$ normalized or rescaled by the void probability $P_0$, i.e.,
\begin{eqnarray}
 Z_A(\vec x) &=& \frac{1}{\Gamma(A+1)} \left[\left(\frac{d}{d z}\right)^A
         Z(\vec x, z)\right]_{z=0}
   = \frac{P_A(\vec x)}{P_0(\vec x)}
\end{eqnarray}
Another type of generating function of $P_A$ that is frequently used
may be defined as
\begin{eqnarray}
 G(\vec x, u) &=& \sum_{A} P_A(\vec x) (1-u)^A
   = \frac{1}{Z(\vec x)} \sum_{A}^\infty Z_A(\vec x) (1-u)^A
         \nonumber  \\
  &=& \frac{Z(\vec x, z=1-u)}{Z(\vec x)}
   = \exp\left[\sum_k x_k [(1-u)^{\alpha_k} - 1] \right]
           \label{genu}   \\
 P_A(\vec x) &=& \frac{1}{\Gamma(A+1)} \left[\left(-\frac{d}{d u}\right)^A
        G(\vec x, u)\right]_{u=1}
\end{eqnarray}
We see that $G(\vec x, 0) = 1$,
$G(\vec x, 1) = P_0(\vec x) = Z_0(\vec x)/Z(\vec x)$
and $G(\vec x, 1-z) = Z(\vec x, z)/Z(\vec x)$.
Once the probability $P_A(\vec x)$ is determined, 
various statistical quantities can be evaluated.
Also the grand canonical partition function $Z(\vec x,z)$ is given
once we know the thermodynamic grand potential 
\begin{eqnarray}
 \Omega(\vec x, z) = - \ln Z(\vec x, z) = - \sum_k x_k z^{\alpha_k}.
\end{eqnarray}
This result can be used to study the statistical properties of a system.
Moreover various moments and cumulants, mean values and fluctuations may
be obtained using the generating function \cite{frag,canon}.

\subsection{Moments and cumulants; 
combinants and hierachical structure}

This subsection gives general expression for various quantities
that will be used later when we discuss specific models.
Since the probability of a specific event $\vec n$ in a grand canonical
ensemble is given by
\begin{eqnarray}
 P(\vec x, \vec n) &=& \frac{W(\vec x, \vec n)}{Z(\vec x)}
\end{eqnarray}
the mean of a quantity $F$ in a grand canonical ensemble is related 
to the mean of $F$ in a canonical ensemble as
\begin{eqnarray}
 <F> &=& \sum_{\vec n} F P(\vec x, \vec n)
   = \frac{1}{Z(\vec x)} \sum_{\vec n} F W(\vec x, \vec n)
   = \sum_A \frac{Z_A(\vec x)}{Z(\vec x)}
        \frac{1}{Z_A(\vec x)} \sum_{\vec n_A} F W(\vec x, \vec n)
              \nonumber  \\
  &=& \sum_A P_A(\vec x) \sum_{\vec n_A} F P_A(\vec x, \vec n)
   = \sum_A P_A(\vec x) <F>_A
\end{eqnarray}
Thus once we obtain $Z_A(\vec x)$ using the recurrence relation
of Eq.(\ref{recur}),
then using the mean in canonical ensemble given by Eq.(\ref{factma}), 
we can obtain mean in grand canonical ensemble.
Since, in a grand canonical enemble, $A$ can be infinity, i.e.,
 $0 \le A < \infty$ with $Z_A(\vec x) = 0$ for $A < 0$
if all $\alpha_k \ge 0$
and $-\infty < A < \infty$ if $\alpha_k$ can be negative,
we have
\begin{eqnarray}
 Z(\vec x) &=& \sum_A Z_A(\vec x)
    = \sum_A Z_{A - m\alpha_k - l \alpha_j}(\vec x)
\end{eqnarray}
for integer $m$ and $l$.
Using this fact, we can see easily that
\begin{eqnarray} 
 <n_k> &=& \sum_{\vec n} n_k P(\vec x, \vec n)
   =\sum_A P_A(\vec x) <n_k>_A
   = \sum_A P_A(\vec x) x_k \frac{Z_{A-\alpha_k}(\vec x)}{Z_A(\vec x)}
              \nonumber  \\
  &=& x_k \frac{\sum_A Z_{A-\alpha_k}(\vec x)}{Z(\vec x)}
   = x_k   \label{meank}     \\
 <M> &=& \sum_{\vec n} \left(\sum_k n_k\right) P(\vec x, \vec n)
   = \sum_k x_k
\end{eqnarray}
This result shows that the weight factor $x_k$ in this model
is the mean number $<n_k>$ in a grand canonical ensemble.
The $m$-th power moment of $A$
and its factorial moments are given simply by
\begin{eqnarray}
 <A^m>(\vec x) &\equiv& \sum_{\vec n}
           \left(\sum_{k=1}^N \alpha_k n_k\right)^m P(\vec x, \vec n)
  =  \frac{1}{Z(\vec x)} \left[\left(z \frac{d}{d z}\right)^m
          Z(\vec x, z)\right]_{z=1}       \\
 \left<\frac{\Gamma(A+1)}{\Gamma(A-m+1)}\right>(\vec x)
  &\equiv& \sum_{\vec n} \frac{\Gamma(A+1)}{\Gamma(A-m+1)} P(\vec x, \vec n)
   = \frac{1}{Z(\vec x)} \left[\left(\frac{d}{d z}\right)^m
          Z(\vec x, z)\right]_{z=1}      
\end{eqnarray}
Similarly the $m$-th cumulants , which is the power moments of $\alpha_k$,
and the factoral cumulants are
\begin{eqnarray}
 <\alpha^m>(\vec x) &\equiv& \left<\sum_{k=1}^N \alpha_k^m n_k\right>
   = \left[\left(z \frac{d}{d z}\right)^m \ln Z(\vec x, z)\right]_{z=1}
   = \sum_{k=1}^\infty \alpha_k^m x_k     \label{alpham}  \\
 f_m(\vec x) &\equiv& \left< \sum_{k=1}^N
          \frac{\Gamma(\alpha_k+1)}{\Gamma(\alpha_k-m+1)} n_k\right>
   = \left[\left(\frac{d}{d z}\right)^m \ln Z(\vec x, z)\right]_{z=1}
              \nonumber   \\
  &=& \left[\left(- \frac{d}{d u}\right)^m \ln G(\vec x, u)\right]_{u=0}
   = \sum_{k=1}^N \frac{\Gamma(\alpha_k+1)}{\Gamma(\alpha_k-m+1)} x_k 
               \label{factfm} 
\end{eqnarray}
We can see easily that, for the power moments of $\alpha_k$,
\begin{eqnarray}
 <\alpha^0> &=& \sum_{k=1}^N x_k = f_0 = <M>    \\
 <\alpha> &=& \sum_{k=1}^N \alpha_k x_k = f_1 = <A>      \label{malph1} \\
 <\alpha^2> &=& \sum_{k=1}^N \alpha_k^2 x_k
           = f_2 + f_1 = <(A - <A>)^2> = <A^2> - <A>^2 = \sigma^2
                        \label{malph2}     \\
 <\alpha^3> &=& \sum_{k=1}^N \alpha_k^3 x_k
           = f_3 + 3 f_2 + f_1 = <(A - <A>)^3>     \label{malph3}    
\end{eqnarray}
The power moments of $\alpha_k$ are directly related to the
power moments of $A$ measured from the mean $<A>$, i.e.,
the cumulants $<\alpha^m>$ are same with the central moments of $A$.
This simple relation does not hold for $m \ge 4$
but we can evaluate them starting from $<\alpha^3>$ 
using the recurrence relation 
\begin{eqnarray}
 <\alpha^{m+1}>(\vec x, z) &=& \left(z \frac{d}{d z}\right)^{m+1} \ln Z(\vec x, z)
   = \left(z \frac{d}{d z}\right) <\alpha^m>(\vec x, z)
\end{eqnarray}
Similarly the $m$-th factorial cumulants $f_m$, which is the factorial moments
of $\alpha_k$, can be found using recurrence relation
\begin{eqnarray}
 \frac{f_{m+1}(\vec x, z)}{f_m(\vec x, z)}
    &=& z \left(\frac{d}{d z}\right) \ln f_m(\vec x, z)  -  m     \label{recfm}
\end{eqnarray}
starting from 
\begin{eqnarray}
 f_0(\vec x, z) &=& <M>(\vec x, z) = \ln Z(\vec x, z) = - \Omega(\vec x, z)
    = \sum_{k=1}^N x_k z^{\alpha_k}    \\
 f_1(\vec x, z) &=& <A>(\vec x, z) = \sum_{k=1}^N \alpha_k x_k z^{\alpha_k}
\end{eqnarray}
The reduced factorial cumulants $\kappa_m$ defined in Ref.\cite{hiera}
corresponds to the factorial cumulants $f_m$ normalized with mean
number $<A> = \bar A$ as
\begin{eqnarray}
 \kappa_m(\vec x, z) &=& \frac{f_m(\vec x, z)}{{\bar A}^m}
\end{eqnarray}
Thus with $\kappa_1 \equiv 1$ and $\kappa_0 = f_0 = <M>$.

A Taylor expansion of
$(p + q)^a$ with a real number $a$, w.r.t. $p$ for $q \ne 0$, is
\begin{eqnarray}
 (p + q)^a &=& \sum_{n=0}^\infty \frac{p^n}{n!}
        \left[\left(\frac{d}{d p}\right)^n (p + q)^a\right]_{p=0}
    = \sum_{n=0}^\infty \frac{\Gamma(a+1)}{n! \Gamma(a-n+1)}
        q^{a-n} p^n
    = \sum_{n=0}^\infty {a \choose n} p^n q^{a-n}   
\end{eqnarray}
Thus, from the factorial structure of $f_m$ and 
from Eqs.(\ref{genu}) and (\ref{factfm}),
\begin{eqnarray}
 G(\vec x, u) &=& \frac{Z(\vec x, z=1-u)}{Z(\vec x)}
   = \exp\left[\sum_{m=1}^\infty \frac{(-u)^m}{m!} f_m(\vec x)\right]
   = \exp\left[\sum_{m=1}^\infty \frac{(-u \bar A)^m}{m!}
          \kappa_m(\vec x)\right]
\end{eqnarray}
Since $P_0(\vec x) = G(\vec x, 1)$
the probability $P_A$ can be obtained by
\begin{eqnarray}
 P_A(\vec x) &=& \frac{1}{\Gamma(A+1)}
      \left[\left(-\frac{d}{d u}\right)^A G(\vec x, u)\right]_{u=1}
   = \frac{(-\bar A)^A}{\Gamma(A+1)}
      \left[\left(\frac{d}{d \bar A}\right)^A P_0(\vec x)\right]
\end{eqnarray}
when $\kappa_m$ is independent of the mean $<A> =\bar A$
as used in Ref.\cite{hiera}.
Using the power moment $<\alpha^m>$ of Eq.(\ref{alpham}),
we also have
\begin{eqnarray}
 G(\vec x, u=1-e^{-\lambda}) &=& \frac{Z(\vec x, z=e^{-\lambda})}{Z(\vec x)}
   = \exp\left[\sum_{m=1}^\infty \frac{(-\lambda)^m}{m!}
          <\alpha^m>(\vec x)\right]
\end{eqnarray}
Since $f_0 = \ln Z = <\alpha^0>$,
\begin{eqnarray}
 Z(\vec x, z) &=& G(\vec x, u=1-z) Z(\vec x) = G(\vec x, u=1-z) e^{f_0(\vec x)}
                  \nonumber   \\
  &=& \exp\left[\sum_{m=0}^\infty \frac{(z-1)^m}{m!} f_m(\vec x)\right]
    = \exp\left[\sum_{m=0}^\infty \frac{(\ln z)^m}{m!} <\alpha^m>(\vec x)\right]
\end{eqnarray}
The generating function $Z(\vec x, z)$ differs from the generating
function $G(\vec x, u)$ only by an extra term of $m = 0$ in their exponent.

The relation between the $x_k$'s and the $Z$ shows that
the $x_k$'s are also the combinants of Ref.\cite{refgyu}.
In turn the combinants $x_k$ can be related to the factorial cumulants $f_m$
defined by
\begin{eqnarray}
 \ln Z(\vec x, z) &=& \sum_k x_k z^{\alpha_k}
   = \sum_{m=0}^\infty \frac{(z-1)^m}{m!} f_m
\end{eqnarray}
The factorial cumulants $f_m$ are the $m$-th order factorial moments 
of $\alpha_k$ of Eq.(\ref{factfm}). Thus  
\begin{eqnarray}
 f_m &=& m! \sum_{k=m}^\infty \pmatrix{k \cr m} x_k
\end{eqnarray}
for $\alpha_k = k$.
The normalized factorial cumulants, i.e., the reduced cumurant, is
\begin{eqnarray}
 \kappa_m &=& f_m/ \bar A^m
    = (m-1)! \kappa_2^{m-1}    \label{kappam}
\end{eqnarray}
for a negative binomial (NB) distribution,
This result of Eq.(\ref{kappam}) shows  that $\kappa_m$ for NB has 
an hierarchical structure of a distribution at the reduced cumulant level
which was realized for the NB distribution in Ref.\cite{hiera}.
This result will be generalized later.

Also using the above power moments and factorial moments we can study
voids and void scaling relation, hierarchical structure, 
combinant and cummulant properties
which will be discussed below.

\subsection{Multi-fragmentation versus multiparticle production}

Since our approach was first used to discuss multifragmentation
and then later extended to include multiparticle production,
we briefly mention some of difference between multifragmentation
and multiparticle production.

In nuclear multifragmentation, $\alpha_k = k$ is the number of nucleons
in a fragment and $n_k$ is the number of fragments of size $k$.
The mean values of the total number of fragments $M = \sum_k n_k$
and the total number of nucleons $A = \sum_k k n_k$ are
\begin{eqnarray}
 <M> &=& \sum_k <n_k> = \sum_k x_k = f_0 = \ln Z    \\
 <A> &=& \sum_k k <n_k> = \sum_k k x_k = f_1   
\end{eqnarray}
In nuclear multifragmentation, the total number of nucleons $A$
is usually fixed. 
In multiparticle production the $A$ is the total number of produced
particles and is not fixed.
Thus in multifragmentation we consider canonical ensemble instead of
a grand canonical ensemble.
The mean multiplicity $<M>_A = \sum_k <n_k>_A$ and 
 $<n_k>_A = x_k Z_{A-k}(\vec x)/Z_A(\vec x)$ in a canonical ensemble
are related with grand canonical ensemble as
\begin{eqnarray}
 <M> &=& \sum_{A=0}^\infty <M>_A P_A    \\
 <n_k> &=& \sum_{A=0}^\infty <n_k>_A P_A = x_k
\end{eqnarray}
The $x_k = <n_k>$ in a grand canonical ensemble.
But the weight factor $x_k$ is different from the mean multiplicity
of cluster size $k$ in a canonical ensemble $<n_k>_A$.
From Eq.(\ref{factma}), $<n_k>_A$ is the weight $x_k$ multiplied
by the ratio of $Z_{A-k}$ and $Z_A$, i.e., $<n_k>_A = x_k Z_{A-k}/Z_A$.
In multifragmentation studies \cite{frag,canon,massd}
we have used $x_k = x z^k/k$ with the same weight $x$ for each cluster
and the same weight $z$ for each nucleon.
This choice of $x_k$ gives
 $Z_A(x,z) = \frac{z^A}{A!} \frac{\Gamma(x+A)}{\Gamma(x)}$ 
and thus $Z_{A-k}/Z_A \to z^{-k}$ as $A \to \infty$ 
where the $z = e^{\beta \mu}$ with $\mu$ the chemical potential.
To determine the weight $x_k$ experimentally, we should use
a grand canonical ensemble, i.e., we need to consider various
system with different values of $A$.

\subsection{Clan parameters and void parameters and void scaling relations}

Van Hove and Giovannini have introduced clan variables $N_c$ and $n_c$ 
to describe a general class of probability distributions,
with most discussions of these variables centering around
the negative binomial distribution \cite{ref12}.  
These variables are defined as
\begin{eqnarray}
 N_c &=& <M> = \ln Z = f_0,  \hspace{1cm}  n_c = <A>/N_c = f_1/f_0
\end{eqnarray}
where the mean number of clans is $N_c$ and the $n_c$ is the mean number of
members per clan.
The $Z$ is the grand canonical generating function 
and thus $N_c = \sum_k x_k$ 
where $x_k$ is the cycle class weight distribution $\vec x$.
The $<A> = \sum_k k <n_k>$ is the mean number of total members (particles).
The $n_k$ here is the number of clans of size $k$ having $k$ members.

The clan variable $N_c$ is also related to the void probability
 $P_0 = Z_0/Z = 1/Z = e^{-N_c}$; thus $N_c = - \ln P_0 = f_0$.
An important function in void analysis is
 $\chi = -\ln P_0/<A> = N_c/<A> = 1/n_c$.
Thus the void parameters, 
void probability $P_0$ and void function $\nu$ \cite{hiera},  
are equivalent to
the generalized clan parameters $N_c$ and $n_c$ 
with the equivalence given by:
\begin{eqnarray}
 f_0(\vec x) &=& \ln Z(\vec x) = -\ln P_0(\vec x) = N_c  \\
 \chi(\vec x) &\equiv& \frac{f_0(\vec x)}{<A>} 
    = \nu = {n_c}^{-1}
\end{eqnarray}
The $- \chi$ is the normalized grand potential $\Omega = -f_0$ 
for the mean $<A>$.

Void analysis looks for scaling properties associated with $\chi$;
specifically, $\chi$ is a function of the combination $<A>\xi$
where $\xi$ is the coefficient of $<A>^2$ in the fluctuation
 $\sigma^2 = <A> + \xi <A>^2$.
Since $f_2 = <\alpha^2> - <\alpha> = <(A - <A>)^2> - <A>$,
the variance of $A$ in a grand canonical ensemble becomes
\begin{eqnarray}
 \sigma^2 &\equiv& <A^2> - <A>^2 =<\alpha^2> = \sum_k \alpha_k^2 x_k
         \nonumber  \\
  &=& <A> + f_2 = <A> + \xi <A>^2 = <A> \left[ 1 + \xi(\vec x) <A>\right]
\end{eqnarray}
with $\xi = \kappa_2$, i.e., the normalized factorial cumulant.  
Since the variance for a Poissonian distribution is the
same as the mean, $\sigma^2 = <A>$, the $\xi = 0$;
thus, the parameter $\xi <A>$ represents a degree of departure from
Poissonian fluctuation normalized by mean $<A>$
of the distribution.
A well known non-Poissonian example is a NB distribution
which has $\xi = \frac{1}{x}$ and 
this becomes Plank distribution with $x = 1$.
Using the recurrence relation Eq.(\ref{recfm}) for $f_m$, we can show that
\begin{eqnarray}
 \xi(\vec x,z) <A>(\vec x) &=& \chi^{-1}(\vec x, z)
      - z \left(\frac{d}{d z}\right) \ln \chi(\vec x, z) - 1
   = \frac{1 - z \left(\frac{d}{d z}\right) \chi(\vec x, z)}
          {\chi(\vec x, z)}  - 1     \\
 \kappa_3(\vec x, z) &=& \frac{f_3(\vec x, z)}{<A>^3}
    = \frac{\xi(\vec x, z)}{<A>} \left(z \frac{d}{d z}\right)
              \ln\left[\xi(\vec x, z) <A>^2\right]
        - 2 \frac{\xi(\vec x, z)}{<A>}  
\end{eqnarray}
A NB distribution has
 $\chi = \ln(1 + \xi<A>)/(\xi<A>)$
while the Lorentz/Catalan (LC) distribution 
discussed in Ref. \cite{prl86} and below
has $\chi = (\sqrt{2\xi<A>+1} - 1)/(\xi<A>)$.
We will study in Sect. \ref{voidscal}, 
$\chi$ {\it vs} $\xi <A>$, i.e., the void or clan variable {\it vs}
the fluctuation for various choices of $x_k$
summarized in Table \ref{tabl1}.

\subsection{Ancestral or evolutionary variables}  \label{ancest}

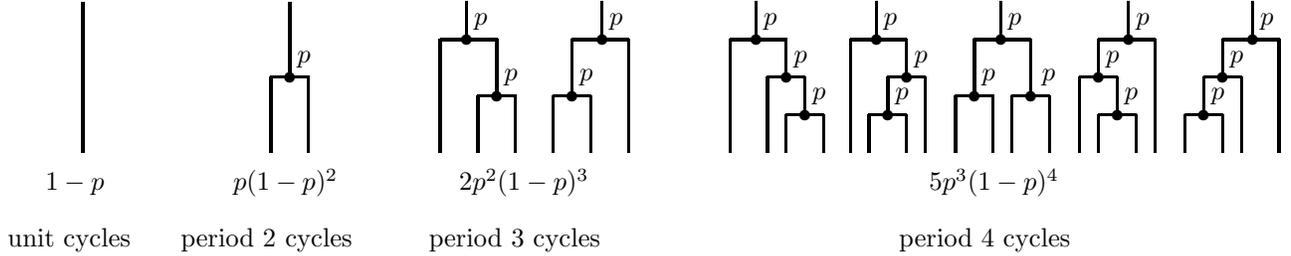
\begin{figure}[hbt]
\setlength{\unitlength}{0.5cm}  
\begin{center}
\begin{picture}(35,7)(1.2,4.0)  
  \thicklines
 \put(3,7){\line(0,1){4}}
    \put(2,6.0){$1-p$}  \put(1,4.5){unit cycles}
 \put(8.5,9){\line(0,1){2}}
    \put(8.0,9){\line(1,0){1}}  \put(8.5,9){\circle*{0.3}}
    \put(8.7,9.4){$p$}
    \put(8.0,7){\line(0,1){2}}  \put(9.0,7){\line(0,1){2}}
    \put(7.0,6.0){$p(1-p)^2$}   \put(5.6,4.5){period 2 cycles}
 \put(13.2,10){\line(0,1){1}}
    \put(12.5,10){\line(1,0){1.5}}  \put(13.2,10){\circle*{0.3}}
    \put(13.4,10.4){$p$}
    \put(12.5,7){\line(0,1){3}}  \put(14.0,8.5){\line(0,1){1.5}}
    \put(13.5,8.5){\line(1,0){1}}  \put(14.0,8.5){\circle*{0.3}}
    \put(14.2,8.9){$p$}
    \put(13.5,7){\line(0,1){1.5}}  \put(14.5,7.0){\line(0,1){1.5}}
   \put(16.8,10){\line(0,1){1}}
    \put(16.0,10){\line(1,0){1.5}}  \put(16.8,10){\circle*{0.3}}
    \put(17.0,10.4){$p$}
    \put(17.5,7){\line(0,1){3}}  \put(16.0,8.5){\line(0,1){1.5}}
    \put(15.5,8.5){\line(1,0){1}}  \put(16.0,8.5){\circle*{0.3}}
    \put(16.2,8.9){$p$}
    \put(15.5,7){\line(0,1){1.5}}  \put(16.5,7.0){\line(0,1){1.5}}
    \put(13.0,6.0){$2p^2(1-p)^3$}   \put(12.2,4.5){period 3 cycles}
 \put(20.9,10){\line(0,1){1}}
    \put(20.2,10){\line(1,0){1.5}}  \put(20.9,10){\circle*{0.3}}
    \put(21.1,10.4){$p$}
    \put(20.2,7){\line(0,1){3}}  \put(21.7,9.0){\line(0,1){1.0}}
    \put(21.2,9.0){\line(1,0){1}}  \put(21.7,9.0){\circle*{0.3}}
    \put(21.9,9.4){$p$}
    \put(21.2,7.0){\line(0,1){2.0}}  \put(22.2,8.0){\line(0,1){1.0}}
    \put(21.7,8.0){\line(1,0){1}}  \put(22.2,8.0){\circle*{0.3}}
    \put(22.4,8.4){$p$}
    \put(21.7,7.0){\line(0,1){1.0}}  \put(22.7,7.0){\line(0,1){1.0}}
  \put(24.1,10){\line(0,1){1}}
    \put(23.4,10){\line(1,0){1.5}}  \put(24.1,10){\circle*{0.3}}
    \put(24.3,10.4){$p$}
    \put(23.4,7){\line(0,1){3}}  \put(24.9,9.0){\line(0,1){1.0}}
    \put(24.4,9.0){\line(1,0){1}}  \put(24.9,9.0){\circle*{0.3}}
    \put(25.1,9.4){$p$}
    \put(24.4,8.0){\line(0,1){1.0}}  \put(25.4,7.0){\line(0,1){2.0}}
    \put(23.9,8.0){\line(1,0){1}}  \put(24.4,8.0){\circle*{0.3}}
    \put(24.6,8.4){$p$}
    \put(23.9,7.0){\line(0,1){1.0}}  \put(24.9,7.0){\line(0,1){1.0}}
  \put(27.4,10){\line(0,1){1}}
    \put(26.7,10){\line(1,0){1.5}}  \put(27.4,10){\circle*{0.3}}
    \put(27.6,10.4){$p$}
    \put(26.7,8.5){\line(0,1){1.5}}  \put(28.2,8.5){\line(0,1){1.5}}
    \put(26.2,8.5){\line(1,0){1}}  \put(26.7,8.5){\circle*{0.3}}
    \put(26.9,8.9){$p$}
    \put(26.2,7){\line(0,1){1.5}}  \put(27.2,7.0){\line(0,1){1.5}}
    \put(27.7,8.5){\line(1,0){1}}  \put(28.2,8.5){\circle*{0.3}}
    \put(28.5,8.9){$p$}
    \put(27.7,7){\line(0,1){1.5}}  \put(28.7,7.0){\line(0,1){1.5}}
   \put(30.8,10){\line(0,1){1}}
    \put(30.0,10){\line(1,0){1.5}}  \put(30.8,10){\circle*{0.3}}
    \put(31.0,10.4){$p$}
    \put(31.5,7){\line(0,1){3}}  \put(30.0,9.0){\line(0,1){1.0}}
    \put(29.5,9.0){\line(1,0){1}}  \put(30.0,9.0){\circle*{0.3}}
    \put(30.2,9.4){$p$}
    \put(29.5,7.0){\line(0,1){2.0}}  \put(30.5,8.0){\line(0,1){1.0}}
    \put(30.0,8.0){\line(1,0){1}}  \put(30.5,8.0){\circle*{0.3}}
    \put(30.7,8.4){$p$}
    \put(30.0,7.0){\line(0,1){1.0}}  \put(31.0,7.0){\line(0,1){1.0}}
   \put(34.1,10){\line(0,1){1}}
    \put(33.3,10){\line(1,0){1.5}}  \put(34.1,10){\circle*{0.3}}
    \put(34.3,10.4){$p$}
    \put(34.8,7){\line(0,1){3}}  \put(33.3,9.0){\line(0,1){1.0}}
    \put(32.8,9.0){\line(1,0){1}}  \put(33.3,9.0){\circle*{0.3}}
    \put(33.5,9.4){$p$}
    \put(32.8,8.0){\line(0,1){1.0}}  \put(33.8,7.0){\line(0,1){2.0}}
    \put(32.3,8.0){\line(1,0){1}}  \put(32.8,8.0){\circle*{0.3}}
    \put(33.0,8.4){$p$}
    \put(32.3,7.0){\line(0,1){1.0}}  \put(33.3,7.0){\line(0,1){1.0}}
   \put(25.5,6.0){$5p^3(1-p)^4$}   \put(24.7,4.5){period 4 cycles}
\end{picture}
\end{center}
\caption{Evolutionary lines of descent in a hierarchical topology.
Each branch increases the cycle length with probability $p$, survival $1-p$.
The probability distribution evolves from Poisson to chaotic.
For clusters each branch generates a bigger cluster.
  }

 \label{LCfig}
\end{figure}

The LC model was shown to be a useful model for discussing
an underlying splitting dynamics when ancestral or evolutionary
variables $p$ and $\beta$ were introduced into $x$ and $z$ as
discussed in Ref.\cite{prl86}.
Percolation or splitting dynamics with
a branching probability $p$ and survival probability $(1-p)$
has a hierarchical topology as shown in Fig.\ref{LCfig}.
Weighting each diagram by $x_k = \beta C_k p^{k-1}(1-p)^k$,
the evolutionary or ancestral variables are related to 
the clan variables $N_c = <M>$ and $n_c = <A>/N_c$.
By taking $C_k$ to be the number of diagrams of size $k$ 
shown in Fig.\ref{LCfig},
the evolutionary dynamics is just that of the LC model.
Then with $\beta$ set equal to 1, $x_1 = (1-p)$,
$x_2 = p (1-p)^2$, $x_3 = 2 p^2 (1-p)^3$, $x_4 = 5 p^3 (1-p)^4$, etc.
The interpretation of this set of $x_k$'s  
reads as follows:
$x_1$ has 1 surviving line without a branch ($p^0(1-p)^1$) and 
one diagram ($C_1 = 1$),
$x_2$ has 1 branch ($p^1$) leading to 2 surviving lines ($(1-p)^2$)
and one diagram ($C_2 = 1$),
$x_3$ has 2 branch points ($p^2$) leading 3 surviving lines ($(1-p)^3$)
and two diagrams ($C_3 = 2$),
$x_4$ has 3 branch points($p^3$), 4 surviving lines ($(1-p)^4$) 
and 5 diagrams ($C_4 = 5$), etc.
In these evolutionary/ancestral variables 
the $f_0$ which determines $Z$ is 
$f_0 = \sum x_k = \beta$ for all $p \le 1/2$.
For $p \ge 1/2$, $f_0 = \sum x_k$ is no longer a constant
and is $f_0 = \sum x_k = \beta (1-p)/p$.

Since the clan variables are $N_c = <M> = f_0$ and $n_c = <A>/N_c = f_1/f_0$,
then, for the LC model with evolutionary or ancestral varables,
\begin{eqnarray}
 N_c &=& <M> = \beta \frac{1 - |1-2p|}{2p}   \\
 n_c &=& <A>/N_c = \frac{2p(1-p)}{|1-2p|(1 - |1-2p|)}
 \\ <A> &=& \beta \frac{(1-p)}{|1-2p|}
 \\  p &=& \frac{1}{2} \left[1 \mp \frac{1}{2 n_c - 1}\right] 
 \\ \beta &=& N_c \frac{n_c - (1 \pm 1)/2}{n_c - 1}
\end{eqnarray}
These can be reduced to $N_c = \beta$, $n_c = (1-p)/(1-2p)$,
and $2p = 1 - 1/(2n_c - 1)$ for $p \le 1/2$
while $N_c = \beta (1-p)/p$, $n_c = p/(2p-1)$, 
and $2p = 1 + 1/(2n_c - 1)$ for $p > 1/2$.
Since the branching probability $p$ varies in $0 \le p \le 1$,
the clan variable $n_c$ has $n_c \ge 1$
with $n_c = 1$ at $p = 0$ and 1 and $n_c = \infty$ at $p = 1/2$.
The behavior of $x_k$ above $p = 1/2$ 
will be discussed in Sec. \ref{sectlc} 
where a $x_0 = \phi_\infty$ will be introduced.
The LC model thus connects the clan variable $n_c$ to the probability $p$ 
of branching in the evolutionary or ancestral picture of Fig.\ref{LCfig} 
or in a percolation model.
For Poisson processes $p=0$ (no branching ), 
$x_k = \beta \delta_{k1}$ (only unit cycles and no BE correlations) and $n_c = 1$
(one member in each clan in average).

\subsection{Various distributions with Gauss hypergeometric series}

In describing either nuclear multifragmentation or multiparticle 
distributions, $\alpha_k$ is taken to be a positive integer $k$.
In a nuclear fragment distribution, an initial number of $A$ nucleons are
clustered into $n_k$ small nuclei of $k$ nucleons.
For multiparticle distributions, $k$ can be related to the length
of the cycle in a cycle class representation of permutations 
for Bosonic or Fermionic symmetry.
Then $A$ is the total number of produced particles of a given type
such as pions and $n_k$ is the number
of cycle classes of cycle length $k$
for a permutation of $A$ identical particles.
These cases have no situation with $k = 0$.
If we consider a formation of $M$ jets which are followed
by a pion creation process, then $k$ may be related
to the number of pions from a jet. 
For this case, $k = 0$ can also be included as
a jet having no pion.

Once we identify an appropriate $x_k$ for a physical system,
then we may use our general model to study the statistical
behavior of the system. 
The various models used in pion distribution can be related
to our general model with $\alpha_k = k$
by choosing $x_k$ as a term in a Gauss hypergeometric
series $F(a, b; c; z)$;
\begin{eqnarray}
 F(a, b; c; z) &=& \sum_{m=0}^\infty \frac{[a]_m [b]_m}{[c]_m} \frac{z^m}{m!}
                \label{hyperf}     \\
 {[a]_m} = \frac{\Gamma(a+m)}{\Gamma(a)}
   &=& \frac{\Gamma(a+n)}{\Gamma(a)} \frac{\Gamma(a+n + m-n)}{\Gamma(a+n)}
           \label{amfc}
   = [a]_n [a+n]_{m-n}       
\end{eqnarray}
Usefull values of $[a]_n$ are
 $[1/2]_n = (2n)!/(n! 2^{2n})$,
 $[1]_n = n!$,
 $[2]_n = (n+1)!$.
Considering only positive $k$, we choose
\begin{eqnarray}
 x_k &=& x \frac{[a]_{k-1} [b]_{k-1}}{[c]_{k-1}} \frac{z^k}{(k-1)!}
      = x \frac{\Gamma(a+k-1)}{\Gamma(a)} \frac{\Gamma(c)}{\Gamma(c+k-1)}
         \frac{\Gamma(b+k-1)}{\Gamma(b)} \frac{z^k}{(k-1)!}
\end{eqnarray}
For this case, the thermodynamic grand potential or the generating 
function is
\begin{eqnarray}
 f_0(\vec x) &=& f_0(x, z) = \log Z(x, z)
   = - \Omega(x, z) = \sum_{k=1}^\infty x_k  
    = x z F(a, b; c; z)
\end{eqnarray}
If we allow jets without a pion, then we may allow $k = 0$ also.
For such a case,
\begin{eqnarray}
 x_k &=& x \frac{[a]_{k} [b]_{k}}{[c]_{k}} \frac{z^k}{k!}
      = x \frac{\Gamma(a+k)}{\Gamma(a)} \frac{\Gamma(c)}{\Gamma(c+k)}
         \frac{\Gamma(b+k)}{\Gamma(b)} \frac{z^k}{k!}   \\
 f_0(\vec x) &=& f_0(x, z) = \log Z(x, z)
   = - \Omega(x, z) = \sum_{k=0}^\infty x_k  
    = x F(a, b; c; z)
\end{eqnarray}
We can see the only difference of the generating functions
between above two cases is the extra factor $z$ for
the grand potential.  
We will mostly concentrate on the first case, i.e., 
$k \ne 0$.

Using Eq. (\ref{factfm}) or the recurrence relation Eq.(\ref{recfm})
and Eqs.(\ref{hyperf}) -- (\ref{amfc}),
\begin{eqnarray}
 f_m(x, z) &=& - z^m \left(\frac{d}{d z}\right)^m \Omega(x, z)
    = z^m \left(\frac{d}{d z}\right)^m \log Z(x, z)
    = z^m \left(\frac{d}{d z}\right)^m xz F(a,b;c;z)   \nonumber \\
   &=& x \frac{[a]_{m} [b]_{m}}{[c]_{m}} z^{m+1}
            F(a+m, b+m; c+m; z)
                \nonumber  \\   & &
     + x m \frac{[a]_{m-1} [b]_{m-1}}{[c]_{m-1}} z^{m}
            F(a+m-1, b+m-1; c+m-1; z)
\end{eqnarray}
The second order normalized factorial cumulant $\xi = \kappa_2$ and
the void variable $\chi$ are then
\begin{eqnarray}
 \xi(x, z) <A> &=& \frac{f_2(x, z)}{<A>}
    = \frac{f_2(x, z)}{f_1(x, z)}        \nonumber  \\
   &=& \frac{\frac{[a]_2 [b]_2}{[c]_2} z^2 F(a+2, b+2; c+2; z)
          + 2 \frac{a b}{c} z F(a+1, b+1; c+1; z)}
        {\frac{a b}{c} z F(a+1, b+1; c+1; z) + F(a, b; c; z)}     \\
 \chi(x, z) &=& \frac{f_0(x, z)}{<A>}
    = \frac{f_0(x, z)}{f_1(x, z)}     \nonumber  \\
   &=& \frac{F(a, b; c; z)}
         {\frac{a b}{c} z F(a+1, b+1; c+1; z) + F(a, b; c; z)}
\end{eqnarray}

For some values of $a$, $b$, and $c$, the hypergeometric
function become a simple function;
\begin{eqnarray}
 F(a, b; b; z) &=& (1-z)^{-a}   \nonumber  \\
 F(a, b; a; z) &=&(1-z)^{-b}            \nonumber   \\
 F(a, 1; 2; z) &=& \frac{1 - (1-z)^{1-a}}{z (1-a)}   \nonumber  \\
 F(1, 1; 2; z) &=& \lim_{a\to 1} F(a, 1; 2; z) = - \frac{\ln(1-z)}{z}
\end{eqnarray}
Various models of pion distributions can be related with
these functions as listed in Table \ref{tabl1}.

\begin{table}[htb]
\caption
{Various models with specific choice of $\alpha_k = k$ and 
$x_k$ in hypergeometric series $F(a, b; c; z)$ of Eq.(\protect\ref{hyperf}).
Here $k=0$ is not included, and
thus $f_0 = \ln Z = \sum_{k=1}^\infty x_k = xz F(a,b;c;z)$.
Here $1 \le k \le N$ 
with $N \to \infty$ except for Poisson which has a finite $Nx$.
 }
    \label{tabl1}
\begin{tabular}{c|c|c|ccc}
\hline
 Model                  &  $x_k$ & $f_0(\vec x) = \ln Z$ & $a$ & $b$ & $c$ \\
\hline
 Poisson (P)            & $N x \delta_{k,1}$ or $x$ for $k = 1, 2, \cdots, N$
     &  $N x = \bar A$  \\
 Geometric (Geo)        &   $x z^k$   &  $\frac{xz}{1-z}$ & $a$  & 1 & $a$  \\
 Negative Binomial (NB) & $\frac{1}{k} x z^k$ & $-x \ln(1-z)$ & 1  & 1 & 2  \\
 Signal/Noise (SN)      & $(y + \frac{x}{k}) z^k$ 
       & $\frac{yz}{1-z} - x\ln(1-z)$ \\
 Lorentz/Catalan (LC) & $\frac{1}{k} 2^{-2(k-1)} \pmatrix{2(k-1) \cr k-1} xz^k$
       & $2x[1 - (1-z)^{1/2}]$  &  $\frac{1}{2}$   &  1  &  2   \\
 Hypergeometric (HGa)   & $\frac{[a]_{k-1}}{k!} x z^k$
      & $\frac{x}{1-a}[1 - (1-z)^{1-a}]$  &  $a$  &  1 &  2  \\
 Random Walk--1d (RW1D) & ${2^{-2(k-1)}} \pmatrix{2(k-1) \cr k-1} x z^k$
      & $ x z (1-z)^{-1/2}$   &  $\frac{1}{2}$  & 1($b$)  & 1($b$)   \\
 Random Walk--2d (RW2D) 
   & $\left[{2^{-2(k-1)}} \pmatrix{2(k-1) \cr k-1}\right]^2 x z^k$
      &  $x z F(\frac{1}{2}, \frac{1}{2}; 1; z)$
        & $\frac{1}{2}$ &  $\frac{1}{2}$  & 1    \\
 Generalized RW1D (GRW1D)
   & $\frac{[a]_{k-1}}{(k-1)!} x z^k$ & $x z (1-z)^{-a}$ & $a$ & $b$ & $b$   \\
 Generalized RW2D (GRW2D)
   & $\left[\frac{[a]_{k-1}}{(k-1)!}\right]^2 x z^k$  & $x z F(a, a; 1; z)$
     & $a$ & $a$ & $1$    \\
  \hline
\end{tabular} 
\end{table}

\begin{table}
\caption{ \protect
Factorial cumulants for various choices of $x_k$ of Table.\protect\ref{tabl1} 
  }  \label{tabl2}
\begin{tabular}{c|c|c|c|c}   
  \hline  
  Model  &  $f_0 = \log Z$  & $f_1 = <A>$  &  $z^{-2} f_2$  &
            $z^{-m} f_m$  \\
    \hline  
 P    & $N x = \bar A$  & $\bar A$  & 0 & 0 for $m \ge 2$ \\
 Geo  & $x \frac{z}{1-z}$  &  $x \frac{z}{(1-z)^2}$   &
      $x \frac{2}{(1-z)^3}$   &  $x \frac{m!}{(1-z)^{m+1}}$           \\
 NB & $- x \ln(1-z)$     &    $x \frac{z}{1-z}$    &
      $x \frac{1}{(1-z)^2}$   &  $x \frac{(m-1)!}{(1-z)^m}$           \\
 SN   & $\frac{yz}{1-z} -x \ln(1-z)$ & $y\frac{z}{(1-z)^2} + x\frac{z}{1-z}$ &
      $y \frac{2}{(1-z)^3} + x \frac{1}{(1-z)^2}$   &
       $\frac{(m-1)!}{(1-z)^m} \left(y \frac{m}{(1-z)} + x\right)$  \\
 LC   & $2 x [1 - (1-z)^{1/2}]$  & $x \frac{z}{(1-z)^{1/2}}$  &
   $x \frac{1/2}{(1-z)^{3/2}}$  &  $x \frac{[1/2]_m}{(1-z)^{m-1/2}}$   \\
 HGa  & $\frac{x}{1-a} [1 - (1-z)^{1-a}]$ &
    $x \frac{z}{(1-z)^a}$  &  $x \frac{a}{(1-z)^{a+1}}$   &
            $x \frac{[a]_{m-1}}{(1-z)^{a+m-1}}$           \\
 RW1D  & $x \frac{z}{(1-z)^{1/2}}$   &
    $\frac{x}{2} \frac{z}{(1-z)^{1/2}} + \frac{x}{2} \frac{z}{(1-z)^{3/2}}$
  & $\frac{x}{4} \frac{1}{(1-z)^{3/2}} + \frac{3}{4} x \frac{1}{(1-z)^{5/2}}$
  & $- x \frac{[-1/2]_m}{(1-z)^{a+1}} + x \frac{[1/2]_m}{(1-z)^{a+2}}$
         \\
 GRW1D   & $x \frac{z}{(1-z)^{a}}$
  & $- x \frac{(a-1) z}{(1-z)^{a}} + {x} \frac{a z}{(1-z)^{a+1}}$
  & $- x \frac{(a-1) a}{(1-z)^{a+1}} + x \frac{a (a+1)}{(1-z)^{a+2}}$
  & $- x \frac{[a-1]_m}{(1-z)^{a+1}} + x \frac{[a]_m}{(1-z)^{a+2}}$
         \\
    \hline  
\end{tabular}
\end{table}

More detail discussions and related physical systems  
of these distributions will be given and discussed in the next section.
Brief discussion about the weight $x_k$ of each model follows.
A Poisson (P) distribution is generated when there is 
monomers only, i.e.,  $x_k = N x \delta_{k,1}$ which gives $f_0 = Nx$.
A Poisson can also be generated if
all the clusters or cycles are treated the same with the same 
weight of $x_k = x$ independent of their size $k$ for $1 \le k \le N$
which also gives $f_0 = Nx$.
The second case can also be viewed as weighting each constituents
by $x^{1/k}$ so that $x_k = (x^{1/k})^k = x$.
The weight $x$ counts the multiplicity $M$
and the multiplicity of clusters (or of monomers for $x_k = Nx \delta_{k,1}$)
has the Poisson distribution.
All the other distributions considered in Table \ref{tabl1} 
have several factors in the weight $x_k$.
One factor in $x_k$ is $z^k$ which is a $k$ dependent geometric term
and comes from assigning the same weight $z$ to each constituents
independent of the cluster or cycle classes it belongs to.
Another factor of weight is independent of $k$ such as the $x$ 
in Table \ref{tabl1} which comes from assigning the same weight $x$
to each cluster or the cycle class as a whole 
independent of its internal structure.
These two factors, $x z^k$, are multiplied by a $k$ dependent or 
independent prefactor.
A geometric (Geo) distribution follows when there is no other weight 
factor beside $x$ and $z$, i.e., no $k$ dependent prefactor 
so that $x_k = x z^k$.
The Geo with $z = 1$ for a finite $N$ is the same as
the Poisson distribution; both have $f_0 = N x$.
The negative binomial (NB) which appears frequently in various
studies has a weight factor assigned to a cluster or cycle
class given by $x z^k/k$.
This has an extra size dependent factor of $1/k$
compared to the geometric distribution.
The signal/noise model (SN) has a two part structure and
interpolates between a Poisson and NB distribution
as will be discussed in Sect. \ref{sectsn}.
The geometric distribution is the signal component of SN
while the NB distribution is the noise component of SN.
The Lorentz/Catalan model (LC) has in its weight a shifted Catalan number
divided by $2^{2(k-1)}$, that is  
$\frac{[1/2]_{k-1}}{k!} = \frac{2^{-2(k-1)}}{k} \pmatrix{2(k-1) \cr k-1}$,
beside the $x z^k$ factor which is the weight for Geo model.
The Catalan numbers given by
 $\pmatrix{2k \cr k}/(k+1)$ are 1, 2, 5, 14, $\cdots$ 
for $k = 1$, 2, 3, 4, $\cdots$ and 
the shifted Catalan numbers given by 
 $\pmatrix{2(k-1) \cr k-1}/k$ are 1, 1, 2, 5,14, $\cdots$.
The importance of this factor is shown in Fig. \ref{LCfig} 
of section \ref{ancest}.  
As can be seen from the arguments of the hypergeometric function
$F(a, b; c; z)$ in Table \ref{tabl1}, the hypergeometric model 
with $b = 1$ and $c = 2$ (HGa) 
include Geo, NB, SN, LC as a special case of HGa 
depending on the value of $a$.
Other models listed in Table \ref{tabl1} are based on random walks.
The use of random walk results was originally due to Feynman \cite{feyn}
in his description of the phase transition in liquid helium.
The random walk aspects arise when considering the closing of cycle
of length $k$. We include them for completeness.
Since the random walk in 1-dimension (RW1D) is the same as LC
except the missing $1/k$ dependence compared to LC,
RW1D can be extended to a generalized RW1D (GRW1D) similar to
the generalization of LC to HGa.
A random walk model in 2-dimension has an extra factor of a shifted
Catalan number and $k 2^{-2(k-1)}$ factor
compared to RW1D and can also be generalized to GRW2D.

Since $k = 0$ is excluded here, 
the partition function for these models are
given simply by a hypergeometric function 
as $Z = \exp[\sum_{k=1}^\infty x_k] = \exp[xz F(a,b;c;z)]$
with various choices of $a$, $b$, $c$.
For example the LC model has $f_0 = \sum_k x_k = x z F(1/2,1;2;z)$
and the NB model has $f_0 = xz F(1,1;2;z)$.
The geometric model $x_k = y z^k$ has
 $f_0 = yz F(a,1;a;z) = yz F(2,1;2;z)$
while the SN model is a combination of the geometric plus NB cases.
These functions are special cases of $f_0 = x z F(a, 1; 2; z)$ of 
the HGa model. 
The generalized random work in 1-dimension (GRW1D) has
$f_0 = xz F(a, b; b; z)$ and
the generaized RW in 2-dimension (GRW2D) has
$f_0 = xz F(a,a; 1; z)$.
The factorial cumulants $f_m$ for these models are summarized
in Table \ref{tabl2}.
The cases with $c = 2$ have a canonical partition function $Z_n$ 
which can be writen in terms of
confluent hypergeometric functions $U(u,v;w)$
and standard factor $z^n/n!$ \cite{prc65}.

\subsection{Generalized model of Hypergeometric (HGa)}  \label{glcbas}

We consider in more detail the hypergeometric model with $b=1$ and $c=2$ (HGa)
here since it includes the NB, Geo, and LC models as special cases.
This generalized model is related with
the hypergeometric function with $b=1$ and $c=2$ 
with an arbitrary value of $a$, 
i.e., $F(a,1;2;z)$, and has the weight of
\begin{eqnarray}
 x_k &=& x z^k \frac{[a]_{k-1}}{k!} = x \frac{z^k}{k!} \frac{\Gamma(a+k)}{\Gamma(a)}
\end{eqnarray}
Its associated grand canonical partition function  
\begin{eqnarray}
 Z(x, z) &=& e^{f_0} = e^{xz F(a,1;2;z)}
    = \exp\left[\frac{x}{(a-1)} \left(\frac{1}{(1-z)^{(a-1)}} - 1\right)\right]    
\end{eqnarray}
is shown in Table \ref{tabl1}.
From Table \ref{tabl2}, we have
\begin{eqnarray}
 f_m(x, z) &=& [a]_{m-1} \frac{x z^m}{(1-z)^{a+m-1}} 
    = x \frac{\Gamma(a+m-1)}{\Gamma(a)} \frac{z^m}{(1-z)^{a+m-1}}   \\
 \kappa_m(x, z) &=& \frac{[a]_{m-1}}{x^{m-1}} (1-z)^{(a-1)(m-1)}
    = \frac{\Gamma(a+m-1)}{\Gamma(a)}
            \left(\frac{(1-z)^{(a-1)}}{x}\right)^{m-1}        \nonumber  \\
  &=& \frac{\Gamma(a + m - 1)}{\Gamma(a)} \frac{\kappa_2^{m-1}(x,z)}{a^{m-1}}
     = A_m \kappa_2^{m-1}(x,z)                \label{kappan}
\end{eqnarray}
The normalized factorial cumulant, i.e., 
the reduced cumulant $\kappa_m$ shows the hierarchical structure
of HGa at the reduced cumulant level 
with $A_m = a^{-(m-1)} \Gamma(a+m-1)/\Gamma(a)$
where $\kappa_m$ is related to $\kappa_2$.
This property was realized for the NB distribution in Ref.\cite{hiera}
which is obtained for Eq.(\ref{kappan}) with $a=1$ giving $A_m = (m-1)!$.
The result of Eq.(\ref{kappan}) is a generalization of the NB result.

Some moments for HGa are
\begin{eqnarray}
 <A>(x, z) &=& f_1(z, \vec x) = \frac{x z}{(1-z)^a}      \\
 \chi(x, z) &=&\frac{f_0(x,z)}{f_1(x,z)}
     = \frac{1}{(1-a)} \frac{(1-z)}{z} \left[(1-z)^{a-1} - 1\right]       \\
 \xi(x, z) &=& \kappa_2(x, z) = \frac{f_2(x,z)}{f_1^2(x,z)} = \frac{a}{x} (1-z)^{(a-1)}  
\end{eqnarray}
Since these relations give
\begin{eqnarray}
 \xi(x, z) <A>(x, z) &=& \kappa_2(x, z) <A>(x, z) 
    =  \frac{f_2(x, z)}{f_1(x, z)}  
    = \frac{a z}{(1-z)} 
\end{eqnarray}
the void parameters can be obtained in terms of the normalized 
fluctuation $\xi$ and the mean number $<A> = \bar A$ by
\begin{eqnarray}
 z(\bar A, \xi) &=& \frac{\xi\bar A}{a + \xi\bar A} = \frac{f_2/\bar A}{a + f_2/\bar A}
    = \frac{f_2}{f_2 + a \bar A}          \\
 x(\bar A, \xi) &=& \frac{\bar A}{z} (1-z)^a
    = \frac{a}{\xi} \left(\frac{a}{a + \xi\bar A}\right)^{(a-1)}
    = \frac{a \bar A^2}{f_2} \left(\frac{a \bar A}{a \bar A + f_2}\right)^{a-1}  \\
 f_0(\bar A, \xi) &=& \log Z(\bar A, \xi) 
  = \frac{x}{a-1} \left[\left(1 + \frac{\xi\bar A}{a}\right)^{a-1} - 1\right]
 = \frac{x}{a-1} \left[\left(1 + \frac{f_2}{a\bar A}\right)^{a-1} - 1\right] \\
 \chi(\bar A, \xi) &=& \frac{f_0}{\bar A}
  =  \frac{1}{(1-a)} \frac{a}{\xi \bar A}
       \left[\left(1 + \frac{\xi \bar A}{a}\right)^{1-a} - 1\right]
   =  \frac{1}{(1-a)} \frac{a \bar A}{f_2}
        \left[\left(1 + \frac{f_2}{a \bar A}\right)^{1-a} - 1\right]
               \label{chiglc}   \\
 \kappa_m(\bar A, \xi) &=& \frac{f_m(\bar A, \xi)}{\bar A ^m}
  = \frac{\Gamma(a+m-1)}{\Gamma(a)} \left(\frac{\xi}{a}\right)^{m-1}
              \label{kapanglc}
\end{eqnarray}
for a given mean value of $<A> = \bar A$ and 
the fluctuation $\xi$ or $\xi\bar A$ or $f_2$.

Table \ref{tabl1} shows that the generalized HGa model becomes 
the Lorentz/Catalan (LC) model with $a = 1/2$,
the negative binomial (NB) model with $a = 1$, 
and geometric (Geo) distribution with $a = 2$.
However the NB should be considered as a $a \to 1$ limit of HGa;  
\begin{eqnarray}
 \lim_{a \to 1} f_0 &=& -x \log (1-z)        \\
 \lim_{a \to 1} Z &=& (1-z)^{-x}             \\
 \lim_{a\to 1} f_0(\bar A, \xi) &=& x \ln \left(1 + \xi\bar A\right)
    = x \ln \left(1 + f_2/\bar A\right)          \\
 \lim_{a \to 1} \chi(\bar A, \xi)
   &=& \frac{1}{\xi \bar A} \log\left(1 + \xi \bar A\right) 
\end{eqnarray}
Thus a NB distribution has
 $\chi = \ln(1 + \xi \bar A)/(\xi \bar A)$ with $x = 1/\xi$
while the LC distribution has $\chi = (\sqrt{1+2\xi \bar A} - 1)/(\xi \bar A)$
with $x = \sqrt{1 + 2 \xi \bar A}/(2\xi)$
and a Geo distribution has $\chi = 2 [1 - (1+\xi \bar A/2)^{-1}]/(\xi \bar A)$
with $1/x = \xi(1 + \xi \bar A/2)/2$
as limiting expressions of a more general $\chi$ given by Eq.(\ref{chiglc}) 
for HGa.

\section{Pion Probability Distribution}

In this section we discuss various features of the pion multipilcity
distribution and its associated fluctuations.
In the introduction we noted the importance of such studies.
The present work is a continuation of the approach developed in 
Ref.\cite{refmek} which is based on Feynman path integral methods \cite{feyn}.
These methods lead to a cycle class decomposition of any permutation,
with permutations appearing when Bose-Einstein or Fermi-Dirac
symmetries are included into the density matrix.
Any permutation of a particle has associated with it a vector $\vec n$
of which $n_k$ is the number of cycles of length $k$.
A cycle of length 2 would be $1 \to 4 \to 1$, etc.
The constraint
 $A = \sum_k k n_k$
must be satisfied for a canonical system.
The $M = \sum_k n_k$ is the multiplicity which varies from 1 to $A$.
A weight $W_A(\vec n, \vec x)$ of Eq.(\ref{weight}) is then given to 
each $\vec n$
with the $x_k$ in $\vec x$ the weight assigned
to a cycle of length $k$.
The grand canonical partition function $Z(\vec x)$ is Eq.(\ref{grandptf})
and the probability distribution $P_n(\vec x)$ of $n = \sum_k k n_k$ pions 
is Eq.(\ref{pax}).  
The mean number $<n>$ and the fluctuation $\sigma^2$ of pions 
can be related to the $x_k$'s as
\begin{eqnarray}
 <n> &=& \sum_k k x_k      \nonumber  \\
 \sigma^2 &=& = <n^2> - <n>^2 = <n> + \xi <n>^2 = \sum_k k^2 x_k
                \label{moment}
\end{eqnarray}
from Eqs.(\ref{malph1}) and (\ref{malph2}).

\subsection{Poisson distribution and cycle of length 2 correlations}

A widely used distribution which also is a basis for comparison
is the Poisson distribution which is obtained for the $x_k$'s by
either having unit cycles only, i.e., $x_k = x_1 \delta_{k1}$
or uniformly distributed cycles independent of their cycle length
which is equivalent to the weight $(x_1/N)^{1/k}$ of each pion in a cycle
class with size $k$ (see Table \ref{tabl1}).  
For this case $n = M$ and 
the cycle class multiplicity $M$ has a Poisson distribution.
Then $<n> = x_1$, $\sigma^2 = <n>$, and $P_n$ is
\begin{eqnarray}
 P_n = <n>^n e^{-<n>}/n!
\end{eqnarray}
with $Z_n = x_1^n / n!$ and $Z = e^{x_1}$.
The probability of no events, the void probability distribution
has $n = 0$ and $P_0 = e^{-<n>}$.
Poisson distributions appear in coherent state emission
and are the Maxwell-Boltzmann limit of Bose-Einstein
and Fermi-Dirac distribution in statistical physics.
Fermi-Dirac and Bose-Einstein statistics lead to departures from
Poisson results.
The voids parameters are $\xi = 0$ and $\chi = 1$ for this distribution.

The first correction to Poisson comes from cycles of length 2.
Then from Eq.(\ref{moment})
$<n> = x_1 + 2 x_2$ and $\sigma^2 = x_1 + 4 x_2 \ne <n>$.
The $P_n$ distribution can be generated for $\vec x = (x_1, x_2, 0, 0, \cdots)$
using the recurrence relation Eq.(\ref{recur}) to obtain $Z_n$ and
 $Z(\vec x) = \exp[x_1 + x_2]$.
The $Z_n$ is simply
\begin{eqnarray}
 Z_n = \frac{1}{n} (x_1 Z_{n-1} + 2 x_2 Z_{n-2})
\end{eqnarray}
The $Z(\vec x) = \sum_n Z_n$ becomes the Hermit polynomial
 $e^{2xz - z^2} = \sum_n \frac{1}{n!} H_n(x) z^n$
when $x_1 = 2xz$ and $x_2 = -z^2$.

A system of one type only gives a Poisson distribution
and a system of two or more types exhibits non-Poissonian behavior.
Departures from Poisson statistics have been noted in level densities 
where random matrix theory gives the Wigner distribution which is
not a Poisson distribution.
The next subsection discusses a simple distribution which is the
negative binomial and it can have large non-Poissonian fluctuations.

\subsection{Negative binomial (NB) distribution}  \label{sectnb}

A distribution frequently discussed for situation that depart 
from Poisson statistics is the negative binomial (NB) disribution.
The cycle class vector $\vec x$ which generates the NB distribution
is $x_k = x z^k / k$, logarithmic distribution, 
and $x_k$ contains cycle lengths $k$ of all sizes.
For the choice (see Tables \ref{tabl1} and \ref{tabl2});
 $Z = (1-z)^{-x}$, $Z_n = \frac{z^n}{n!} \Gamma(x+n)/\Gamma(x)$,
 $<n> = xz/(1-z)$, $\sigma^2 = xz/(1-z)^2 = <n>(1 + <n>/x)$.
The $P_n$ can be written in the well-known form
\begin{eqnarray}
 P_n &=& \frac{z^n}{n!} (1-z)^x \frac{\Gamma(x+n)}{\Gamma(x)}
   = \pmatrix{n+x-1 \cr n} \left(\frac{1}{1 + \frac{<n>}{x}}\right)^x
        \left(\frac{<n>/x}{1 + <n>/x}\right)^n
\end{eqnarray}
The $x=1$ limit is also referred to as a Planck distribution
since the fluctuation is $\sigma^2 = <n> (1 + <n>)$
and has the characteristic Planck behavior.
Departures from Poisson statistics arise from chaotic sources
as discussed in quantum optics.
Intermittency and its association with chaos have also been
studied using the negative binomial distribution \cite{ref9}.
The voids parameters are
 $\xi = \frac{1}{x}$ and $\chi = \frac{\ln(1 + \xi <n>)}{\xi <n>}$
for this distribution.

Ref. \cite{ref10} gives several sources for the origin of the 
negative binomial distribution.
These sources include sequential processes that arise from a composite
of logarithmic and Poisson distributions,
self-similar cascade processes and connections with Cantor sets
and fractal structure, generalization of the Plank distribution,
solutions to stochastic differential equations.
Becattini etal \cite{becat} have shown that the NB distribution arises
from decaying resonances.
The $\alpha$-model of Ref. \cite{ref9}, which is a self-similar
random cascading process, leads to NB like behavior.
The stochastics aspects of the NB distribution have been discussed
by R. Hwa \cite{rhwa}.
Hegyi \cite{hegyi} has discussed the NB distribution in terms of combinants.

\subsection{Signal/noise (SN) model; Coherent and chaotic emission}
    \label{sectsn}

The signal/noise model (SN in Table \ref{tabl1} and \ref{tabl2}) 
offers a continuous connection between a Poisson distribution ($x=0$ 
with $y \to 0$ and $z \to 1$ keeping $<n>=yz/(1-z)^2$ finite) 
and the Planck or Bose-Einstein (BE) ($x=1$ and $y=0$) or 
more generally a negative binomial distribution (arbitrary $x$ with $y=0$).
Since a coherent state is a Poisson emitter and a Planck
or NB distribution comes from a chaotic state,
the SN model interpolates between coherent and chaotic emission.
The SN model is discussed in Ref.\cite{poistrn} and its connection
to the cycle class picture developed here was initially shown
in Ref.\cite{prl86}.
Here, further developments of it are discussed.
To begin with the cycle weights have a geometric piece $y z^k$,
distributed with $z^k$,
and a NB piece $x z^k/k$ or $x_k = (y +(x/k)) z^k$.
The $y$ and $z$ can be redefined in terms of a coherent signal (Poisson emitter)
variable $S$ and a thermal Bose-Einstein noise variable $N$ using 
 $y = \frac{S}{(N/x)(1+N/x)}$ and $z = \frac{N/x}{(1+N/x)}$.
The probability distribution is
\begin{eqnarray}
 P_n &=& \frac{z^n}{Z} L_n^{x-1}(-y)
    = \frac{(N/x)^n}{(1+N/x)^{x+n}} \exp\left[\frac{- S}{1 + N/x}\right]
                         L_n^{x-1}(\frac{-S}{(N/x)(1+N/x)})
\end{eqnarray}
with $x =1$ being the Glauber-Lach model
and $L_n^a$ is an associated Laguerre function.   
When $N \to 0$ the Poisson limit is realized with $x_k = S \delta_{k,1}$
and when $S \to 0$, the distribution goes into a NB distribution.
In terms of $S$ and $N$ the mean
\begin{eqnarray}
 <n> &=& \frac{yz}{(1-z)^2} + \frac{xz}{1-z}= S + N
\end{eqnarray}
while the fluctuation
\begin{eqnarray}
 \sigma^2 &=& <n> + \xi <n>^2
    = \frac{yz}{(1-z)^2} + \frac{xz}{1-z} + \frac{xz^2}{(1-z)^2} + \frac{2 y z^2}{(1-z)^3}
         \nonumber  \\ 
   &=& S + N + \frac{N^2}{x} + \frac{2 S N}{x}
\end{eqnarray}
The void parameters are
\begin{eqnarray}
 \xi &=& \frac{f_2}{f_1^2} = \frac{N}{x} \frac{(2S+N)}{(S+N)^2}     \\
 \chi &=& \frac{f_0}{f_1} = \frac{ x \ln(1+N/x) + S/(1+N/x)}{S+N}
\end{eqnarray}
for this distribution.
The SN model has important application to quantum optics
and, in particluar, to photon counts from lasers \cite{qoptic}.
Biyajima \cite{biyaj} has suggested using it for particle
multiplicity distribution as does Ref. \cite{poistrn}.
When the noise level $N \to 0$, $\xi \to 0$ and $\chi \to 1$.
When the signal level $S \to 0$, $\xi \to 1/x$ 
and $\chi \to \frac{x}{N} \ln(1 + N/x)$.

\subsection{Lorentz/Catalan (LC) distribution; underlying splitting
probability}     \label{sectlc}

Probability distribution associated from field emission from
line shapes which are Lorentzian have appeared in quantum optics.
Its cycle class description was developed in Ref.\cite{prl86}
and is $x_k = x C_k z^k / 2^{2(k-1)}$
with $C_k = \pmatrix{2(k-1) \cr k-1}/k$ a shifted Catalan number,
distributed with NB with $x=1/2$ (see Eq.(\ref{xkpknb}) in Sect. \ref{secseq}).
The partition function is $Z(x,z) = \exp(2x[1 - (1-z)^{1/2}])$ and
the probability distribution associated with it is
\begin{eqnarray}
 P_n(x, z) &=& e^{-2x (1 - \sqrt{1-z})} (2x)^{2n} U(n, 2n; 4x) \frac{z^n}{n!}
\end{eqnarray}
with $U(u,v;w)$ a confluent hypergeometric function.
The void parameters are  $\xi = \frac{1}{2x \sqrt{1-z}}$
and $\chi = \frac{1}{\xi <n>}\left[\sqrt{1 + 2 \xi <n>} - 1\right]$
with $z = 2 \xi <n> /(1 + 2 \xi <n>)$
for this distribution.

The results of quantum optics, in the notation of Ref. \cite{qoptic},
can be obtained \cite{prl86} when $x = T \Omega/2$,
 $z = 2 W \gamma/\Omega^2$, $\Omega^2 = \gamma^2 + 2 W \gamma$.
The $W$ is an integral of the Lorentzian line shape
 $\Gamma(\omega) = b/[(\omega - \omega_0)^2 + \gamma^2]$,
 $T$ is the time, and $2 x \sqrt{1-z} = \gamma T$.

An interesting feature of the model arises when evolutionary
or ancestral variables are introduced which contain a 
branching probability $p$, survival probability $(1-p)$ 
as discussed in Sect.\ref{ancest}.
Percolation or splitting dynamics with evolutionary parameters $p$ 
and $\beta$ become
equivalent to the LC model with clan variables $N_c = \beta$ 
and $n_c = \frac{(1-p)}{(1-2p)}$ for $p < 1/2$.
Specifically $x = \beta/4p$ and $z = 4p(1-p)$ giving
$x_k = \beta C_k p^{k-1} (1-p)^k$.
In these evolutionary/ancestral variables 
the $f_0 = \sum x_k = 2 x (1 - \sqrt{1-z})$ which determines $Z$ has the
interesting property $f_0 = \sum x_k = \beta$ for all $p \le 1/2$.
For $p \ge 1/2$, $f_0 = \sum x_k$ is no longer a constant
and is $f_0 = \sum x_k = \beta (1-p)/p$.
To keep $f_0 = \sum_k x_k$ a constant 
without changing $<n> = \sum_k k <n_k> = \beta \frac{(1-p)}{|1-2p|}$,
a $x_0 = \phi_\infty$ was introduced in Ref.\cite{prl86}.
The $\phi_\infty = 0$ for $p \le 1/2$ 
and $\phi_\infty = \beta (2p-1)/p$ for $p \ge 1/2$.
For $p \ge 1/2$ there is a finite probability that the splitting
will go on forever.   
In percolation above a certain $p$ an infinite cluster is formed and 
$\phi_\infty$ is similar to the strength of the infinite cluster.
Moreover the sudden appearance of $\phi_\infty$ is similar
to the behavior of an order parameter in a phase transition.
The appearance of $\phi_\infty$ can also be interpreted as
the sudden appearance of a jet without pion ($k = 0$), 
when $x_0 = \phi_\infty$.

A connection of the LC model can also be made with a Ginzberg-Landau
theory of phase transitions and a Feynman-Wilson Gas.
These connections are discussed in Ref. \cite{prc65}.

\subsection{Thermal models}

Thermal emission of pions based on statistical mechanics
and equilibrium ideas have been popular descriptions of pions
coming from relativistic heavy ion collisions.
For thermal models \cite{refmek}, the
 $x_k = (V T^3/2\pi^2) (m/T)^2 K_2(km/T)/k$
for a cycle length $k$ or a cluster of size $k$
with $K_2$ a Mac Donald function.
For low temperatures, 
 $x_k = (V/\lambda_T^3) e^{-km/T} /k^{5/2}$ and the 
Boltzmann factor in mass, $e^{-km/T}$, suppresses large fluctuations.
In the high temperature limit and/or zero pion mass limit
 $x_k = (V/\pi^2) T^3 /k^4$.
The $x_k$ can be used to generate the pion probability distribution $P_n$.
The thermal models can be combined with hydrodynamic descriptions
and an application was given \cite{refmek} to $158 A$ GeV Pb+Pb data measured
by the CERN/NA44 and CERN/NA49 collaborations.
The results of Ref.\cite{refmek} showed a Gaussian distribution
with a width about 20 \% larger than the Poisson result.
For Table II of Ref.\cite{refmek} has $x_1 = 260$, $x_2 = 9.957$,
$x_3 = 1.097$, $x_4 = 0.163$, $\cdots$.
The $x_k$'s have a rather sharp fall off with $k$ and two terms $x_1$ 
and $x_2$ give a reasonably good description of the Gaussian.
A Poisson emitter with large $<n>$ has also a Gaussian distribution
with a width such that $\sigma^2 = <n>$.   
For the first two $x_k$'s just given $x_1 = 260$, $x_2 = 10$,
the $\sum x_k = 270$, $<n> = \sum k x_k = 280$
and $\sigma^2 = 300$.

\subsection{Fisher exponent $\tau$ and other distrubutions}

Various forms for $x_k$ encountered for the cycle class weights
have a form $x_k = x z^k/k^\tau$.
The NB has $\tau = 1$, the geometric has $\tau = 0$
and the SN has a combination.
The LC model has $\tau = 3/2$ for large $k$ using Stirling approximation.
In clusters yields $\tau$ is called Fisher critical exponent
and it determines the droplet yields $<n_k>$ around a critical point
of a first order liquid-gas phase transition.
It has been studied extensively in medium energy heavy ion 
collisions \cite{advnp}.
Here we discuss its importance in particle multiplicity distributions.
A power law of $<n_k>$ is the power law of $x_k$ according to
Eq.(\ref{meank}) in grand canonical ensemble
and Eq.(\ref{factma}) in canonical ensemble with finite size effect
through $Z_A$.
The $\tau$ determines the grand canonical partition function
 $Z(\vec x) = \exp[\sum_k x_k] = \exp[x \sum_k z^k /k^\tau]$.
Bose-Einstein condensation of atoms in a box of sides $L$
of dimensions $d$ have $x = L^d/\lambda_T^d$
with $\lambda_T = h/(2\pi m k_B T)^{1/2}$
and have $\tau = 1 + d/2$.
Feynman used random walk arguments, the closing of a cycle parallels
a closed random walk, to discuss his choice of $x_k$ in his discussion
of a superfluid phase transition in liquid equilibrium.
We have therefore consider two other case for $x_k$ which
asymptotically have $\tau = 1/2$ and $\tau = 1$.
These are called random walk (RW) in 1D and 2D respectively.
The RW1D has an $x_k$ which is just $k$ times the $x_k$ of the LC model.
The RW2D has the same $\tau$ as the NB distribution in the
asymptotic limit of large $k$.

\subsection{Pion laser model}

A simple emitting source model for pions was introduced by
Pratt \cite{ref7} and solved analytically by
Cs\"{o}rg\"{o} and Zim\'{a}ny \cite{ref7a}.
The model can be used to study the role of Bose-Einstein
symmetrization effects on pionic distributions.
An intersting property of this model is that a Poisson emitter
of pions with strength $\eta$ can become a pion ``laser''.
A manifestation of this property is a large enhancement of pions
in a zero momentum mode.
For this model the 
 $x_k = (\eta/\eta_c)^k /k/(1-(\gamma_ - /\gamma_+)^{k/2})^3$
with $\gamma_\pm = \frac{1}{2} (1 + y \pm \sqrt{1+y})$,
$y = 2 R^2 m T - 1/2$ and $\eta_c = \gamma_+^{3/2}$.
For very large $k$, $x_k = z^k/k$ \cite{refmek},
a NB behavior with $x = 1$ or the Planck limit. The $z = \eta/\eta_c$.
For small $k$ and typical values of $R$ and $T$
the $x_k = (R^2 m T)^{3/2} (\eta/(R^2mT)^{3/2})^k/k^4$.
The $1/k^4$ dependence also appears in thermal models with zero mass pions.

\section{Comparison of Probability distributions within HGa}

The hypergeometric  
case (HGa) which we studied more in detail in Sect. \ref{glcbas}
included various cases of Table \ref{tabl1}; Geo, NB, LC
which are distinguished by one parameter $a$ of the hypergeometric model.
Thus we use HGa to compare various models for pion distribution
and other particle distributions.

\subsection{Voids and void scaling relations}  \label{voidscal}

Void analysis looks for scaling propoerties associated with $\chi = f_0/<n>$;
specifically, $\chi$ is a function of the combination $\xi<n> = f_2/<n>$
where $\xi = \kappa_2$ is the coefficient of $<n>^2$ in the fluctuation
 $\sigma^2 = <n^2> - <n>^2 = <n> + \xi <n>^2$.
For example, a NB distribution has
 $\chi = \ln(1 + \xi<n>)/(\xi<n>)$
while the LC distribution has $\chi = (\sqrt{2\xi<n>+1} - 1)/(\xi<n>)$.
These two cases are limiting expressions of a more general $\chi$ given by
\begin{eqnarray}
 \chi = \frac{\left(1 + \xi <n>/a\right)^{1-a} - 1}{(1-a)\xi<n>/a}
\end{eqnarray}
of the HGa model.
The NB $\chi$ is obtained from the limit $a \to 1$
while the LC $\chi$ follows for $a = 1/2$.

\begin{figure}[htb]
\centerline{
   \epsfxsize=4.0in   \epsfbox{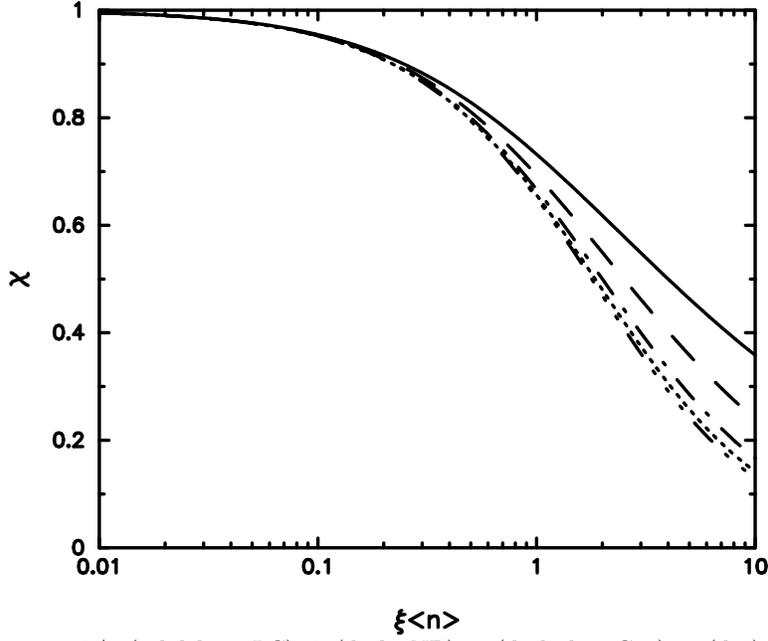}
}
\caption{$\chi$ vs $\xi <n>$ for $a = 1/2$ (solid line; LC), 
1 (dash; NB), 2 (dash-dot; Geo), 3 (dot), 4 (dash-dot-dot-dot).
For Poisson distribution $\xi = 0$ and $\chi=1$.}
 \label{hierafig}
\end{figure}

We show the void variable $\chi$ as a function of $\xi <n>$ in Fig.\ref{hierafig}.
This shows that we can vary $a$ to fit data.
Ref.\cite{hiera} claims the NB distribution fits reasonably well 
the void distribution for single jet events in $e^+ e^-$ annihilation 
but Fig.\ref{hierafig} shows 
all of the curves might fit such data up to $\xi<n> \sim 1$
since the various curves are reasonably close up to this $\xi <n>$.
Thus further investigations of this data is required to distinguish
various models. 
Higher moments or cumulants might need to be compared for this purpose.
Within HGa, $\kappa_3(\bar n, \xi, a) = \left(\frac{a+1}{a}\right) \xi^2$ 
for a given value of $<n> = \bar n$ and $\xi$ 
according to Eq.(\ref{kapanglc}).
Thus $A_3 = \kappa_3/\xi^2 = (a+1)/a = 3$, 2, 3/2, 4/3, 5/4 
for $a = 1/2$, 1, 2, 3, 4 independent of $\xi$.
The $\kappa_3$ may help in distinguishing various models
for the data with a small value of $\xi$, i.e., the region of $\xi \bar n < 1$
in Fig.\ref{hierafig}.
The differences of $A_m$ between models with different value of $a$ 
becomes larger as the order $m$ of the reduced factorial cummulant
becomes higher.

\subsection{Probability distribution}

\begin{figure}[htb]
\centerline{
   \epsfxsize=5.5in   \epsfbox{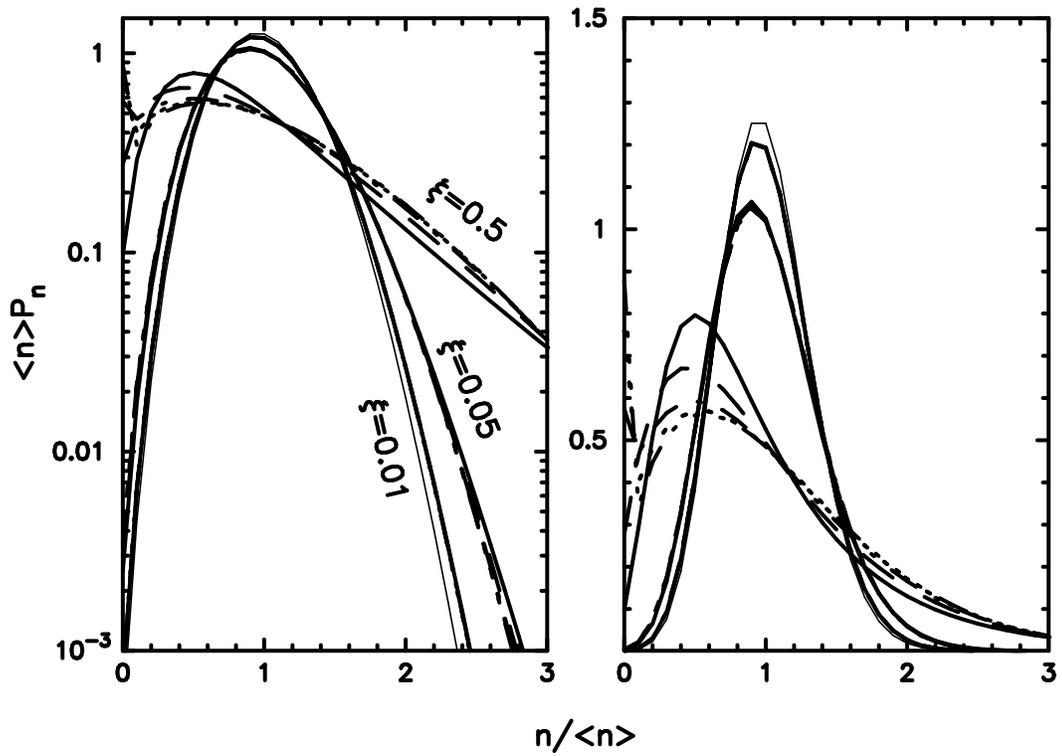}
}
\caption{$P_n$ for fixed $<n> = 10$ for $a = 1/2$ (solid line), 1 (dash), 
2 (dash-dot), 3 (dot), 4 (dash-dot-dot-dot)
and for $\xi = 0.01$, 0.05, and 0.5
in log scale on the left and linear scale on the right.
For $\xi = 0.01$ all distributions become
very close to Poisson (thin solid curve) already.
($P_0$ becomes large for large $\xi$.)}
 \label{cankno}
\end{figure}

\begin{figure}[htb]
\centerline{
   \epsfxsize=5.5in   \epsfbox{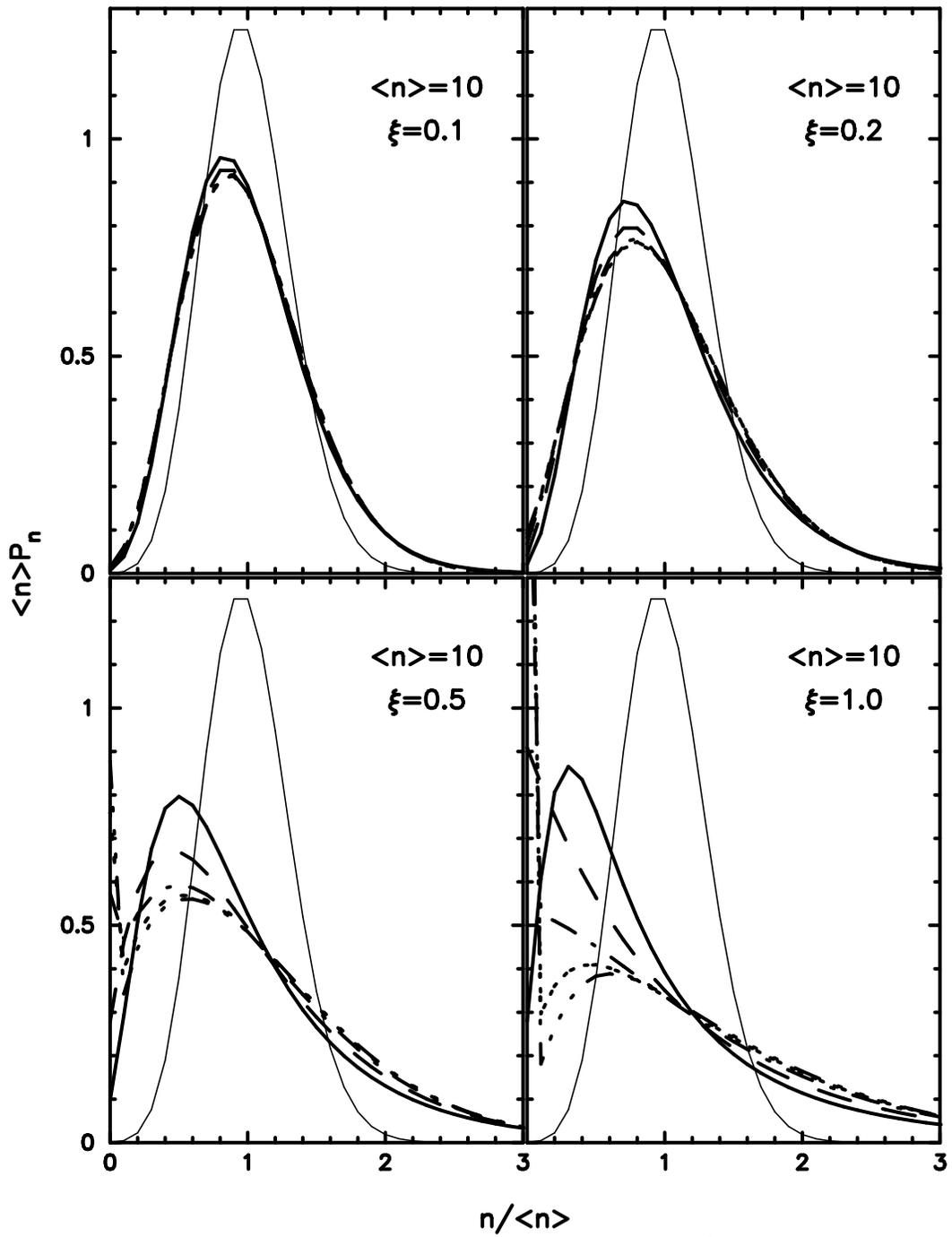}
}
\caption{ 
$P_n$ for fixed $<n> = 10$ and for $a = 1/2$, 1, 2, 3, 4. 
The value $\xi = f_2/<n>^2$ are shown in each figure.
The various choices of $a$ for each curve are given in the figure
caption of Figs. \protect \ref{hierafig} and \protect \ref{cankno}.   }
 \label{canpna}
\end{figure}

\begin{figure}[htb]
\centerline{
   \epsfxsize=5.5in   \epsfbox{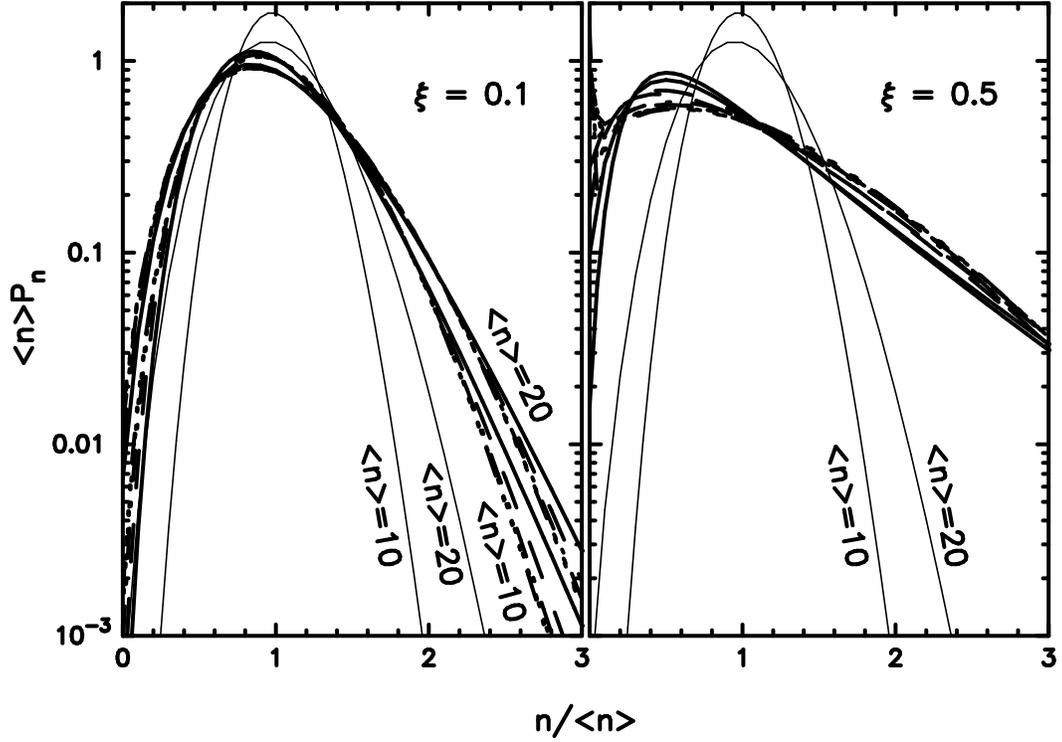}
}
\caption{KNO plot of $<n>P_n$ for fixed $<n> = 10$ and 20 for $a = 1/2$, 1, 2, 3, 4. 
 }
 \label{knoscl}
\end{figure}

Once we know the generating function $Z(\vec x)$ or $f_0(\vec x)$
we may study the probability distribution  $P_n$ 
using Eq.(\ref{pax}) or the recurrence relation Eq.(\ref{recur});
\begin{eqnarray}
 P_n(x, z) &=& \frac{Z_n(x, z)}{Z(x, z)}
         = \frac{1}{Z(x,z)} \frac{z^n}{n!}
            \left[\left(\frac{d}{d z}\right)^n Z(x,z)\right]_{z=0}
\end{eqnarray}
for $n = \sum_k k n_k$.
For NB which is a special case of HGa in the $a \to 1$ limit,
 $x_k = x \frac{z^k}{k}$,
\begin{eqnarray}
 Z^{\rm NB}(x,z) &=& (1 - z)^{-x} = P_0^{-1}(x, z)    \\   
 P_n^{\rm NB}(x,z) &=& \frac{1}{Z} \frac{z^n}{n!} \frac{\Gamma(x+n)}{\Gamma(x)}
    = (1-z)^x \frac{z^n}{n!} \frac{\Gamma(x+n)}{\Gamma(x)}
             \label{nbprob}
\end{eqnarray}
The $P_n$ for various cases of Table \ref{tabl1} are shown
in Fig.\ref{cankno} and Fig.\ref{canpna} with the same mean value 
 $<n> = \bar n$ and the same fluctuation $\xi = \kappa_2 = f_2/\bar n^2$
for fixed $<n> = 10$.
Fig.\ref{knoscl} shows KNO plots of $<n> P_n$ versus $n/<n>$ 
for fixed $<n> = 10$ and 20.
Fig.\ref{cankno} shows that the various models considered here
have almost the same distribution for small fluctuation ($\xi = 0.01$)
and in this case they are very similar to a Poisson's distribution.
For $\xi = 0.05$ the models are similar to each other except 
for larger $n/\bar n$
even though they are different from a Poisson distribution.
For larger fluctuations such as $\xi = 0.5$, the models have very different
forms even if they have the same mean value and fluctuation.
Fig.\ref{canpna} shows that the probability distribution of these models 
differ in their form for fluctuations larger than $\xi \approx 0.2$.

The KNO behavior for $\bar n \to \infty$ can also be studied 
for the models listed in Table \ref{tabl1}.
According to KNO scaling, for distributions with a large mean value,
the distribution $<n> P_n$ becomes model independent in the new 
variable $n/\bar n$, i.e., variable scaled by mean value.
In general, any distribution becomes Gaussian for large mean $<n>$
according to the central limit theorem; specifically
\begin{eqnarray}
 P_n(\bar n, \sigma) &=& \frac{1}{\sqrt{2\pi\sigma^2}}
               \exp\left[-\frac{(n - \bar n)^2}{2\sigma^2}\right]   \nonumber  \\
   &=& \left(1 + \frac{1}{\xi \bar n}\right)^{-1/2}
          \frac{1}{\bar n \sqrt{2\pi\xi}}
               \exp\left[-\frac{1}{2\xi}\left(\frac{n}{\bar n} - 1\right)^2
              \left(1 + \frac{1}{\xi \bar n}\right)^{-1} \right] 
       \label{gauskno}
\end{eqnarray}
with the mean $<n> = \bar n$ and the standard deviation $\sigma$
which is related to the reduced factorial cumulant $\xi$ by
 $\sigma^2 = <n^2> - <n>^2 = \bar n + \xi {\bar n}^2$.
This means that the KNO scaling follows when the fluctuation is
given by $\sigma^2 = \xi {\bar n}^2$ with a constant $\xi$
or when $\xi \bar n \gg 1$.
Thus to compare KNO scaling properties of $\bar n P_n$ for different models,
the fluctuation of these models should have
the same value of $\xi = f_2/\bar n^2$.
For a small $\xi$, the Poisson component of the fluctuation
 $\sigma^2 = \bar n + \xi {\bar n}^2$
becomes dominant and thus the KNO scaling behavior is broken.
For large $\xi$ the Poisson component is negligible and 
KNO scaling is realized, i.e., 
\begin{eqnarray}
 \bar n P_n(\bar n, \xi) &=& \frac{1}{\sqrt{2\pi\xi}}
        \exp\left[-\frac{1}{2\xi}\left(\frac{n}{\bar n} - 1\right)^2\right]
     \label{knolim}
\end{eqnarray}
The KNO plot of Fig.\ref{knoscl} shows that the various distributions 
have no KNO scaling property for small fluctuation ($\xi = 0.1$)
but show a KNO scaling property for large fluctuation ($\xi = 0.5$).
The effect of mean on $\bar n P_n$ is larger than the difference 
between different model for $\xi = 0.1$ 
while the effect of mean on $\bar n P_n$ is much smaller than the difference 
between different models exhibiting KNO scaling behavior for $\xi = 0.5$.
Ref.\cite{poistrn} shows the total charge distribution in hadronic collisions.
From their fits we can extract the corresponding fluctuation which
are $\xi = 0.05 \sim 0.5$ and $\bar n = 6 \sim 13$.
This means that the KNO scaling behavior is marginal for these data, i.e.,
just fitting the distribution $\bar n P_n$ of the data does not show clear 
evidence for KNO scaling.
We need to evaluate the explicit values of the mean number $\bar n$ and
the fluctuation $\xi$ to check the KNO behavior of this data;
the value of $\xi \bar n$ should be large enough to show KNO scaling
behavior as can be seen from Eqs.(\ref{gauskno}) and (\ref{knolim}).

\subsection{Sequential procedures and compound Poisson distributions}
    \label{secseq}

A Poisson distribution plays a very important role in physics.
As already noted, in statistical physics, Maxwell-Boltzmann statistics 
leads to Poisson probabilities.
Other distributions are compared to the Poisson distribution
which acts as a benchmark for comparison.
The distributions considered in this paper can have large 
non-Poissonian fluctuations.
The purpose of this section is to show how they can be rewritten as
a compound process or sequential process involving one
aspect that has a Poisson character.
As an example the negative binomial distribution can be obtained
from a compound Poisson-logarithmic distribution as discussed
in Ref.\cite{poistrn}.
Here, we extended this result to include the other distributions
and we also show that the final distribution can be obtained
from compounding it with another distribution,
such as the negative binomial.
In general, the underlying picture for a sequential process
involves a two step procedure in which the observed particles
arise from the production of ``clusters'' with the
subsequent decay of each cluster producing its distribution
of particles.
The final distribution is obtained by compounding the 
probability distribution of the clusters with another distribution
coming from each cluster and suming over clusters.

The LC model was shown to be a useful model for discussing
an underlying splitting dynamics when ancestral or evolutionary
variables $p$ and $\beta$ where introduced into $x$ and $z$ as
shown in Sects. \ref{ancest} and \ref{sectlc} 
and discussed in Ref.\cite{prl86}.
The LC model thus connects the average size of clan $n_c$ 
to the probability $p$ of branching in the ancestral picture of Fig.\ref{LCfig}.
For $p < 1/2$, the mean number of clans is $N_c = \beta$ and
the mean number of members per clan is $n_c = (1-p)/(1-2p)$.
For Poisson processes $p=0$, $x_k = \beta \delta_{k1}$
(only unit cycles and no BE correlations) and $n_c = 1$.

Further discussions of the mean number of members per clan or of $n_c$
will now be given.
The $x_k = <n_k>$ with $<n_k>$ the mean number of cycles
or correlation of size $k$ (c.f., Eq.(\ref{meank})).
The $n_k$ can also be considered as the number of clusters of size $k$.
This is a natural identification when applying this approach
to nuclear multifragmentation.
Then the mean number of clans $N_c$ is the mean multiplicity of clusters
$<M> = \sum_k <n_k>$  and the mean number of members per clan
\begin{eqnarray}
 n_c &=&  \frac{<n>}{N_c} = \frac{<n>}{<M>}
\end{eqnarray}
is the mean size of clusters.
The use of a underlying cluster picture to describe the negative
binomial multiplicity distribution can be found in 
Ref.\cite{poistrn}.
In such a picture the observed particles arise from production 
of $M = c$ clusters with probability distribution $P_c$.
This is sequentially followed by
each cluster decaying into $k_\alpha$ particles
with the probability $P_{k_\alpha}$
with $\alpha = 1$, 2, $\cdots$, $c$.
The probability of observing $n = \sum_k k n_k = \sum_\alpha k_\alpha$ 
particles is then obtained by a compound probability expression
\begin{eqnarray}
 P_n = \sum_c \sum_{\{k_\alpha\}} P_c \prod_{\alpha=1}^c P_{k_\alpha}
\end{eqnarray}
A negative binomial distribution can be obtained when
 $P_c = <c>^c e^{-<c>} / c!$
and $P_{k_\alpha} = (q^{k_\alpha} / k_\alpha) / \ln(1/p)$.
Here $c = M$ and $<c> = <M>$, $p = (1-z) = (1+<n>/x)^{-1}$
and $q = 1-p = z = (<n>/x) / (1+<n>/x)$.
Also $N_c = <c> = x \ln(1 + <n>/x)$ and $n_c = <n>/N_c = (<n>/x) / \ln(1 + <n>/x)$.
This structure can be generalized as follow.

Since the generator of Poisson distribution is an exponential, i.e.,
the expansion of exponetial gives the Poisson's distribution
 $P_n^{\rm P}(\bar n)$, \begin{eqnarray}
 e^{\cal N} &=& \sum_{M=0}^\infty \frac{{\cal N}^M}{M!}
      = e^{\cal N} \sum_M e^{-{\cal N}} \frac{{\cal N}^M}{M!}
      = e^{\cal N} \sum_{M=0}^\infty P_M^{\rm P}({\cal N})
\end{eqnarray}
The grand partition function or the generating function $Z = e^{f_0}$ 
for any distribution can be represented as a Poisson distribution whose
mean value is the void variable $f_0$ or the grand potential $\Omega = - f_0$.
On the other hand and in general we can rewrite the void variable as
\begin{eqnarray}
 f_0(\vec x) &=& \ln Z(\vec x) = \sum_{k} x_k
     = {\cal N} \sum_{k} {\cal P}_k(\vec x)
\end{eqnarray}
where ${\cal N} = f_0 = \sum_k x_k$ and
\begin{eqnarray}
 {\cal P}_k(\vec x) &=& \frac{x_k}{\cal N}    
\end{eqnarray}
The ${\cal P}_k(\vec x)$ can be connected to its generating 
function ${\cal G}(\vec x, u)$:
\begin{eqnarray}
 {{\cal G}(\vec x, u)} &=& \sum_k (1-u)^k {\cal P}_k(\vec x)
    = \frac{1}{\cal N} \sum_k x_k (1-u)^k
\end{eqnarray}
Thus the generating function $G(\vec x, u)$ of $P_n$ can be expanded
in terms of ${\cal P}_k$ as
\begin{eqnarray}
 G(\vec x, u) &=& \sum_n (1 - u)^n P_n(\vec x)
      = \frac{1}{Z(\vec x)} \sum_n Z_n(\vec x) (1 - u)^n
     = \frac{e^{{\cal N G}(\vec x, u)}}{e^{\cal N}}
                      \nonumber \\
    &=& \sum_{M=0}^\infty \frac{1}{e^{\cal N}} \frac{[\cal N G]^M}{M!}
     = \sum_M P_M^{\rm P}({\cal N}) \left[\sum_j (1-u)^j {\cal P}_j\right]^M
                      \nonumber \\
     &=& \sum_M P_M^{\rm P}({\cal N})
               \sum_{\vec n_M} \prod_k [(1-u)^k {\cal P}_k]^{n_k}
\end{eqnarray}
where $P_M^{\rm P}({\cal N}) = e^{-{\cal N}} {\cal N}^M/M!$ is
the Poisson distribution with the mean of ${\cal N} = f_0$.
Here the sum over $\vec n_M$ is the sum over partitions $\vec n$
with a fixed $M = \sum_k n_k$.
Thus we have
\begin{eqnarray}
 P_n(\vec x) &=& \sum_M P_M^{\rm P}({\cal N})
          \sum_{\vec n_M} \prod_k {\cal P}_k^{n_k}(\vec x)   \label{comprb}
\end{eqnarray}
with $M = \sum_k n_k$ and $n = \sum_k k n_k$.
Any distribution obtained from a generating function of the form
of $Z(\vec x) = e^{f_0} = \exp[{\sum_k x_k}]$
can therefore be decomposed as a compound Poisson's distribution
with some other distribution ${\cal P}_k = x_k/f_0$ obtained from
the weight $x_k$. 

The sequential nature of a process is explicitly shown on Eq.(\ref{comprb}).
The observed particle multiplicity distribution arises from a
two step process in which $M = \sum_k n_k$ clusters are first 
distributed according to a Poisson distribution.
This is then sequentially followed by breaking each
of the $n_k$ clusters of type $k$ into $k$ particles
with probability ${\cal P}_k = x_k/{\cal N}$ and with $n = \sum_k k n_k$.
The probability associated with a given $M$ and $\vec n$ with $\vec x$
is $P_M(\vec x, \vec n) = P_M^{\rm P}({\cal N}) \prod_k {\cal P}_k^{n_k}
    = P_M^{\rm P}({\cal N}) \prod_k (x_k/{\cal N})^{n_k}$.

As an illustration we consider the LC model 
with $x_k = x C_k z^k/2^{2(k-1)}$.
Using the evolutionary variables \cite{prl86} $x = \beta/4p$ and $z = 4p(1-p)$
then ${\cal N} = \sum x_k = \beta$ for $p \le 1/2$ as already noted
so that $P_M^{\rm P} = e^{-\beta} \beta^M/M!$.
The $x_k = \beta C_k p^{k-1} (1-p)^k$ so that
 ${\cal P}_k = x_k/{\cal N} = C_k p^{k-1} (1-p)^k$.
The underlying diagram associated with ${\cal P}_k$ are shown 
in Fig.\ref{LCfig}.
For a negative binomial (NB) distribution, $x_k = x z^k/k$ and 
thus ${\cal P}_k = x_k/{\cal N}$ is generated
from ${\cal N} = \sum_k x_k = - x \ln (1-z)$.
Therefore the NB is a compound Poisson-Logarithmic
distribution as shown in Table \ref{tabl3}.

As another example, 
we consider the HGa model with general $a$ instead of $a = 1/2$ for LC 
or $a = 1$ for NB.
The weight $x_k$ has the structure of the probability $P_k(x,z)$ of NB 
distribution given by Eq.(\ref{nbprob}), i.e.,
\begin{eqnarray}
 x_k &=& x \frac{z^k}{k!} \frac{\Gamma(a+k-1)}{\Gamma(a)}
    = \frac{x}{a-1} \frac{1}{(1-z)^{a-1}} P_k^{\rm NB}(a-1,z)     \label{xkpknb}
\end{eqnarray}
thus ${\cal P}_k = [1 - (1-z)^{a-1}]^{-1} P_k^{\rm NB}$ for HGa.
Therefore the HGa  $P_n$ distribution is a compound Poisson-NB distribution.
This may interpreted as a sequential process in which clusters with
a Poisson cluster distribution $P_c$ 
breakup into particles with a particle distribution ${\cal P}_k$
given by a NB distribution.
For the various models considered here with their $x_k$ listed 
in Table \ref{tabl1},
the corresponding distribution ${\cal P}_k$ and the normalization 
factor ${\cal N} = f_0$ are listed in Table \ref{tabl3}.
We can further see that HGa can be looked as a 
compound Poisson-Poisson-Logarithmic distribution, i.e., 
a distribution having three sequential steps;
Poissonian breakup into clusters $\to$ Poissonian breakup of each 
cluster $\to$ logarithmic breakup of each of them.

\begin{table}[htb]
\caption{  
Poissonian sequential distribution for various models 
of Table \protect \ref{tabl1}.
 }    \label{tabl3}
\begin{tabular}{c|c|c|l} 
 \hline 
  Model  &  Weight ${\cal N}$ & Distribution $\ {\cal P}_k$
       &  Comments on ${\cal P}_k$  \\
 \hline 
  P    & $<n>$ & $\frac{1}{N}$ 
       & Monomer only or Uniform for $N$ species  \\  
  Geo  &  $x \frac{z}{1-z}$  & $(1-z) z^{k-1}$ 
       &  Uniform with constituents   \\
  NB     & $x \ln\left(\frac{1}{1-z}\right)$    &
           $\frac{z^k}{k} / \ln\left(\frac{1}{1-z}\right)$
       &  logarithmic with constituents  \\
  LC   & $2x \left[1 - \sqrt{1-z}\right]$  &
      $\left[\frac{1/2}{1 - \sqrt{1-z}}\right]
               \frac{z^k}{k!} \frac{\Gamma(k-1/2)}{\Gamma(1/2)}$
       &  NB with constituents with ${\cal P}_0 = 0$   \\
  HGa & $\frac{x}{1-a} \left[1 - (1-z)^{1-a}\right]$  &
     $\left[\frac{(1-z)^{1-a}}{(1-z)^{1-a} - 1}\right] P_k^{\rm NB}$ 
       &  NB with constituents without $k = 0$   \\
 $ x_k = \frac{x}{k!} z^k$  &  $x e^{a z}$  &  $e^{-a z} \frac{z^k}{k!}$
       &  Poisson (exponential)    \\
 \hline     
\end{tabular}    
\end{table}

Because of a unique role played by the Poisson distribution
and the form $e^{f_0} = \exp[{\sum_k x_k}]$ of the generating function,
the cluster distribution $P_c$ is usually taken to be a Poisson.
However, as noted before, other divisions are possible.
Using the same approach used above for $P_c = P_M^{\rm P}$,
we can expand any distribution
using a NB instead of Poisson, i.e., 
with $P_c = P_M^{\rm NB}$  using
the form of the generating function $Z(x,z) = (1-z)^{-x}$ for the NB.
Replacing $z$ by ${\cal N}(z)$, 
the normalization factor of a new distribution, we have
\begin{eqnarray}
 Z(x,z) &=& \left[1 - {\cal N}(z)\right]^{-x}
     = \left[1 - {\cal N}(z) {\cal G}(u=0)\right]^{-x}
\end{eqnarray}
The ${\cal G}(u)$ is
\begin{eqnarray}
 {\cal G}(u) &=& \frac{{\cal N}([1-u]z)}{{\cal N}(z)}
    = \sum_k (1-u)^k {\cal P}_k
\end{eqnarray}
while the $G(u)$ is
\begin{eqnarray}
 G(u) &=& \sum_n (1 - u)^n P_n
    = \frac{Z(x, [1-u]z)}{Z(x, z)}
    = \frac{\left[1 - {\cal N} {\cal G}(u)\right]^{-x}}
           {\left[1 - {\cal N}\right]^{-x}}
\end{eqnarray}
Thus we have
\begin{eqnarray}
 G(u) &=& \sum_{M=0}^\infty (1 - {\cal N})^x \frac{{\cal N}^M}{M!}
         \frac{\Gamma(x+M)}{\Gamma(x)} [{\cal G}(u)]^M
    = \sum_{M=0}^\infty P_M^{\rm NB}(x, {\cal N}) [{\cal G}(u)]^M
         \nonumber  \\
  &=& \sum_{M=0}^\infty P_M^{\rm NB} 
             \left[\sum_{j=0}^\infty (1-u)^j {\cal P}_j\right]^M 
   = \sum_{M=0}^\infty P_M^{\rm NB}
        \sum_{\{n_k\}_M}  \prod_{k} \left[(1-u)^k{\cal P}_k\right]^{n_k}     \\
 P_n(\vec x) &=& \sum_M P_M^{\rm NB}(x, {\cal N}) 
        \sum_{\{n_k\}_M} \prod_k {\cal P}_k^{n_k}(\vec x)   \label{compnb}
\end{eqnarray}
The result of Eq.(\ref{compnb}) shows that
the distribution $P_n$ can be written as a compound probability distribution
of a negative binomial $P_M^{\rm NB}$ with another probability ${\cal P}_k$
distribution generated from ${\cal G}(u)$.
For the case of ${\cal N}(z) = e^z$, 
which may be considered as the fugacity $e^z = e^\mu$ for a particle
with chemical potential $\mu = z$,
${\cal G}$ can further be decomposed as
\begin{eqnarray}
 {\cal G}(u) &=& \frac{{\cal N}((1-u)z)}{{\cal N}(z)} = \frac{e^{(1-u)z}}{e^z}
    = \sum_{k=0}^\infty (1-u)^k P_k^{\rm P}(z)
\end{eqnarray}
i.e., ${\cal P}_k$ for this case is Poisson $P_k^{\rm P}(z)$.
If ${\cal N} = 1 - e^{f_0/x}$, then ${\cal P}_k = P_k^{\rm P}(f_0/x)$
without $k = 0$ and $Z(x,z) = \left[1 - (1-e^{f_0/x})\right]^x = e^{f_0}$.
For $f_0 = \sum_k x_k$ given in Table \ref{tabl1},
the ${\cal P}_k$ becomes the same probability with $x_k$ replaced
by $x_k/x$. As an example the HGa with
 $Z(x,z) = \exp\left(\frac{x}{1-a}[1 -(1-z)^{1-a}]\right)$
can be decomposed as a sequential process consisting of a NB 
distribution of clusters with $Z(x,{\cal N}=1-e^{f_0/x})$ 
which is then followed by a breakup of clusters distributed with
a HGa distribution 
with $Z(x=1,z)$ but without voids, i.e., with ${\cal P}_0 = 0$.
Thus this decomposition separates the parameter $x$ assigned to cluster 
from other parameters.

\subsection{Poisson transformation and other transformation}

Compound distribution such as those considered in the previous section
can also be understood using
the fact that the Laplace transform of $G(u)$ is related to the
Poisson transform for the probability distribution $P_n$ \cite{poistrn};
\begin{eqnarray}
 G(u \bar n) &=& \int_0^\infty d y f(y) e^{-u y \bar n}
        = \int_0^\infty d y f(y) G^{\rm P}(-u y \bar n)        
\end{eqnarray}
where $G^{\rm P}(\bar n) = e^{-\bar n} e^{(1-u) \bar n}$ is 
the generating function for Poisson's distribution and
\begin{eqnarray}
 \int_0^\infty d y f(y) &=& 1     \nonumber  \\
 \int_0^\infty d y y f(y) &=& 1          \\
 \int_0^\infty d y y^m f(y) &=& {\bar n}^{-m} \left<\frac{n!}{(n-m)!}\right>
                    \nonumber
\end{eqnarray}
For a Poisson distribution,
\begin{eqnarray}
 f^{\rm P}(y) &=& \delta(y - 1)     \\
 P_n^{\rm P}(\bar n) &=& \frac{(\bar n)^n}{n!} e^{-\bar n}   \\
 G^{\rm P}(u \bar n) &=& \sum_{n=0}^\infty (1-u)^n P_n^{\rm P}(\bar n)
   = \sum_{n=0}^\infty \frac{\left[(1-u) \bar n \right]^n}{n!}
         e^{- \bar n}
   = e^{-u \bar n}
\end{eqnarray}
By a Laplace transform or Poisson transform,
\begin{eqnarray}
 G(u \bar n) &=& \sum_{n=0}^\infty (1-u)^n P_n(\bar n)
    = \int_0^\infty dy f(y) e^{-u y \bar n}
    = \int_0^\infty dy f(y) G^{\rm P}(u y \bar n)
                   \nonumber   \\
   &=& \sum_{n=0}^\infty (1-u)^n
        \int_0^\infty dy f(y) \frac{(y \bar n)^n}{n!} e^{-y \bar n}  
    = \sum_{n=0}^\infty (1-u)^n \int_0^\infty dy f(y) P_n^{\rm P}(y \bar n)       \\
 P_n(\bar n) &=& \int_0^\infty dy f(y) \frac{(y \bar n)^n}{n!} e^{-y \bar n}
     = \int_0^\infty dy f(y) P_n^{\rm P}(y \bar n)   
\end{eqnarray}
where $P_n^{\rm P}(y \bar n)$ is the Poissonian probability 
with a mean of $y \bar n$.
The Poisson transform of $f(y)$ is $P_n(\bar n)$ and
the Laplace transform of $f(y)$ is $G(u \bar n)$.
Thus $f(y)$ is an inverse Poisson transform of $P_n(\bar n)$ or 
an inverse Laplace transform of $G(u \bar n)$.
However the probability $P_n(\bar n)$ which is a Poisson transform of $f(y)$ 
may also be considered as a Laplace transform of
$f(y) \frac{(y \bar n)^n}{n!}$ instead of $f(y)$ itself.
The probability $P_n(\bar n)$ is a superposition of a Poisson distribution
$P_M^{\rm P}(y \bar n)$ with weight of $f(y)$.
Thus we may interprete the $P_n(\bar n)$ with 
a mean number $\bar n$ as the probability
in an ensemble of mixed systems with various values of
the energy or temperature which is distributed with weight $f(y)$.
Each of the system with a fixed energy or temperature
breakups to give a Poisson distribution $P_n^{\rm P}(y \bar n)$
with a scaled mean number $y \bar n$.  

Since the $G(z,u) = Z((1-u)z)/Z(z)$,
we have $G(z,u=1) = Z_0(z)/Z(z) = Z^{-1}(z)$ for the case of $Z_0 = 1$
and thus
\begin{eqnarray}
 \left[\frac{1}{Z(z)}\right] &=&
    \int_0^\infty dy f(y) e^{-y \bar n}
     = \int_0^\infty dy f(y)   
         \left[\frac{1}{Z^{\rm P}(y\bar n)}\right]
\end{eqnarray}
Thus $f(y)$ is also an inverse Laplace transform of $1/Z(z)$ 
with $u = 1$, i.e., $1/Z^{\rm P}(y\bar n) = e^{-y\bar n}$.
In general, the Laplace transform and its inversion are
\begin{eqnarray}
 F(s) &=& \int_0^\infty e^{-s t} f(t) dt     \nonumber \\      
 f(t) &=& \frac{1}{2\pi i} \int_{c-i\infty}^{c+i\infty} e^{st} F(s) ds
\end{eqnarray}
Some examples are 
\begin{equation}   \label{laplac}  
\begin{array}{c@{\hspace{33pt}}c@{\hspace{33pt}}l}
 \hline
   F(s)  &  f(t)  &  {\rm constraint \ and \ comments} \\
 \hline
 \frac{1}{(s+a)^n}  &  \frac{t^{n-1} e^{-at}}{(n-1)!}
         &   (n=1,2,3,\cdots) \ \ \ \ \ \ {\rm NB} \\
 e^{-k\sqrt{s}} & \frac{k}{2\sqrt{\pi t^3}} \exp\left[- \frac{k^2}{4t}\right]
         &   (k > 0)  \ \ \ \ \ \ {\rm LC} \\
    \hline
\end{array}      
\end{equation}
For NB with $Z^{\rm NB}(x,z) = (1 - z)^{-x}$ and $\bar n = \frac{xz}{(1-z)}$,
\begin{eqnarray}
 G(x,z,u) &=& \frac{(1-z)^x}{(1 - (1-u)z)^x} = \left(1 + u \frac{z}{1-z}\right)^{-x}
  = \frac{x^x}{(u \bar n + x)^x}      \\
 Z^{-1}(x,z) &=& (1-z)^x = \frac{x^x}{(x + \bar n)^x} = G(x,z,1)     
\end{eqnarray}
Thus $P^{\rm NB}$ is  the Poisson transform of
\begin{eqnarray}
 f(y) &=& \frac{x^x y^{x-1} e^{-xy}}{(x-1)!} = x \frac{(xy)^{x-1} e^{-xy}}{(x-1)!}
\end{eqnarray}
which is the inverse Laplace transform of $n^n (s+a)^{-n}$
with $s = \bar n$ and $a = x$ and $n = x$ from Eq.(\ref{laplac})
as shown in Ref.\cite{poistrn}.
With $x = 1$, the NB becomes Bose-Einstein (BE) distribution 
with $f(y) = e^{-y}$.
For LC with $Z^{\rm LC}(x,z) = e^{2x(1 - \sqrt{1-z})}$
and $\bar n = \frac{xz}{\sqrt{(1-z)}}$,
\begin{eqnarray}
 G(x,z,u) &=& \exp\left[2x\sqrt{1-z} \left(1 - \sqrt{1 + \frac{u}{x\sqrt{1-z}}
        \frac{xz}{\sqrt{1-z}}}\right)\right]    \nonumber  \\
  &=& e^{2x\sqrt{1-z}} \exp\left[-2\sqrt{x\sqrt{1-z}}
           \sqrt{x\sqrt{1-z} + u \bar n}\right]  \\
 Z^{-1}(x,z) &=& \exp\left[-2x(1 - \sqrt{1-z})\right]
    = \exp\left[2x\sqrt{1-z}
      \left(1 - \sqrt{\frac{{1 - z + z}}{{1-z}}}\right)\right]     \nonumber   \\
  &=& e^{2x\sqrt{1-z}} \exp\left[-2\sqrt{x\sqrt{1-z}}
           \sqrt{x\sqrt{1-z} + \bar n}\right]  = G(x,z,1)
\end{eqnarray}
Thus $P^{\rm LC}$ is the Poisson transform of
\begin{eqnarray}
 f(y) &=& e^{2x \sqrt{1-z}}   
      \frac{\sqrt{x\sqrt{1-z}}}{\sqrt{\pi y^3}}
         \exp\left[-\frac{x\sqrt{1-z}}{y} - y x\sqrt{1-z}\right] 
\end{eqnarray}
which is the inverse Laplace transform of $e^{-k\sqrt{s}} e^{2x \sqrt{1-z}}$
with $s = \bar n + x\sqrt{1-z}$ and $k = 2 \sqrt{x\sqrt{1-z}}$
from Eq.(\ref{laplac}).

Similarly we can also decompose any distribution as a superposition
of NB distribution instead of superposition of Poisson distribution
by replacing $G^{\rm P}(y \bar n) = e^{-y \bar n}$ in Laplace
transform by the generating function
 $G^{\rm NB}(x,z,u) = (1 + u \frac{z}{1-z})^{-x} = \left(\frac{x}{x + u \bar n}\right)^x$ 
of NB distribution.
That is using Mellin transform
 $F(s) = \int_0^\infty f(y) y^{s-1} dy$,
\begin{eqnarray}
 G^{\rm NB}(u \bar n, x) &=& \sum_{n=0}^\infty (1 - u)^n P_n^{\rm NB}(\bar n, x)
    = \frac{x^x}{(x + u \bar n)^{x}}
    = x^x \left(u \bar n + x\right)^{(-x+1) - 1}
            \\
 G(u \bar n, x) &=& \sum_{n=0}^\infty (1 - u)^n P_n(\bar n, x)
    = \int_0^\infty f(y')  x^x (u y' \bar n)^{(-x + 1) - 1} dy'
                \nonumber  \\
   &=& \int_0^\infty f(y + x/u\bar n)   
        x^x (u y \bar n + x)^{(-x + 1) - 1} dy
                \nonumber  \\
   &=&  \int_0^\infty f(y + x/u\bar n) G^{\rm NB}(u y \bar n, x) dy   \\
 P_n(\bar n, x) &=& \int_0^\infty f(y + x/u\bar n)  P_n^{\rm NB}(y \bar n, x) dy
\end{eqnarray}
with $s = 1-x$.
Thus a NB transform for the probability $P_n$ may be defined 
as a shifted Mellin transform for the generator $G(u)$.
Further investigation of these transformation properties
may give interesting features of various distributions.
As a special case of Mellin transform
 $(s-1)! = \int_0^\infty e^{-t} t^{s-1} dt$.

\section{conclusion}

Event-by-event studies from ultrarelativistic heavy ion collisions
such as done at RHIC are being used to study the details of
particle multiplicity distributions as, for example,
those associated with pions.
Such studies not only give information about the mean number of
particles produced, but also information about fluctuations
and higher order moments of the probability distribution
which are important tools for studying the underlying processes
and mechanisms that operate.
They are also useful in distinguishing various phenomenological models.
Issues associated with fluctuations play an important role in
many areas of physics and departures from Poisson statistics are 
of current interest.
One purpose of this paper was an investigation of various models of
particle multiplicity distributions that can be used in
event-by-event analysis.
This study was done using a generalized model based on a
hypergeometric series (HGa) 
and uses a grand canonical ensemble as its basic framework.
This framework has its origin in a Feynman path integral approach
and involves a cycle class decomposition of the permutation
symmetries that originate from the underlying wavefunctions associated
with the produced particles.
Many of the existing distributions used in particle phenomenology are
shown to be special cases of this more general hypergeometric model.
Several quantities are shown to be quite general;  
such as heirarchical scaling relations (Eq.(\ref{kappan})),
initially discused in terms of a particular distribution such as
the negative binomial distribution.  
Various models and associated distribution can be developed
in a unified way.
These include the Poisson distribution coming from coherent emission,
chaotic emission producing a negative binomial distribution,
combinations of coherent and chaotic process leading to signal/noise
distributions and field emission from Lorentzian line shapes
producing the Lorentz/Catalan distribution which are all shown
to be special cases of the HGa model.
The HGa model and its associated special cases are used to explore 
a wide variety of phenomena.
These include; linked pair approximations leading to heirarchical
scaling relations on the reduced cumulant level,
generalized void scaling relations,
clan variable descriptions and their connections with stochastic
variables associated with branching processes,
KNO scaling behavior, enhanced non-Poissonian fluctuations.

Our results show that
even though various distributions have the same mean and fluctuation
the distribution itself or the underlying mechanism could be very different. 
Thus to find the correct distribution and underlying mechanism from
the data more informations than just the mean and its fluctuation
are necessary.
The model used here, HGa, has three parameters, $x$, $z$, and $a$.
Thus with a given value of $a$, there is no extra controlling parameter 
for a given mean and fluctuation.
All the higher moments and cumulants are then determined by the model
without any further controling parameter.
Further studies of general models having more parameters 
are needed where the higher moments can be used as extra conditions
in comparing various distributions.

We compared various pionic distribution within a generalized
hypergeometric model (HGa) which is a special case of much more
general distribution in grand canonical ensemble.
In our model used here for a grand canonical ensemble, 
the weight factor $x_k$ for each
species is the same as the mean number $<n_k>$ of the species
in the grand canonical ensemble. Thus we may determine
the weight $x_k$ through experimental data. 
This result also shows that any power law behavior in $<n_k>$
is directly related to the power law of $x_k$.

It is explicitly shown that KNO scaling works only for large fluctuations
for all the distribution related to the HGa model.
Comparison within the HGa model also shows that
just comparing void variables $\chi$ and $\xi$  is not enough
to distinguish different models that describing pion data.
Thus new parameters have to be found which are quite different
between different models.
Beside mean values and fluctuations
higher order reduced factorial cumulants need also to
be evaluated.

In this paper we have also generalized the Poisson transformation and 
the compound distribution that arises from sequential process.
Specifically, the underlying sequential picture involves a two step process
where the final distribution arises from the production of clusters
followed by a subsequent decay of the clusters.
For the HGa model, the final distribution is obtained from 
compounding a Poisson distribution of clusters with a NB distribution
coming from the decay of each of the clusters.
The HGa may arise through a three step sequential process of
Poisson-Poisson-Logarithmic compound distribution.
It is also shown that the HGa can arise from
a two step sequential process of a NB distribution followed by a new HGa 
with a different mean value.

This work was supported in part by Grant No. 2001-1-11100-005-3 
from the Basic Research Program of the Korea Science and
Engineering Foundation
and in part by the DOE Grant No. DE-FG02-96ER-40987.


\begin{thebibliography}{99}

\bibitem{ref1} NA49 Collaboration, G. Roland et al,
    Nucl. Phys. {\bf A638}, 91c (1998).
\bibitem{ref2} NA49 Collaboration, H. Appelh\"{a}user etal,
    Phys. Lett. {\bf B459}, 679 (1999).
\bibitem{ref3} K. Rajagopal and F. Wilczek, Nucl. Phys. {\bf B404}, 557 (1993).
\bibitem{ref4} S. Gavin and B. M\"{u}ller, Phys. Lett. {\bf B329}, 486 (1994).
\bibitem{ref5} G. Bayer and H. Heiselberg, Phys. Lett. {\bf B469}, 7 (1999).
\bibitem{ref6} V. Heinz and B.V. Jacak, Ann. Rev. Nucl. Part. Sci., {\bf 49},
      (1999).
\bibitem{ref7} S. Pratt, Phys. Lett. {\bf B301}, 159 (1993).
\bibitem{ref7a} Cs\"{o}rg\"{o} and J. Zimanyi, Phys. Rev. Lett. {\bf 80},
    916 (1998).
\bibitem{ref8} H.R. Andrews, C.G. Townsend, H.J. Meisner, D.S. Durfee,
   D.M. Korn, W. Ketterle, Science {\bf 275}, 637 (1997).
\bibitem{ref9} A. Bialas and R. Peschanski, Nucl. Phys. {\bf B272}, 703 (1986);
   {\bf B306}, 857 (1988).
\bibitem{ref10} P. Carruthers and C.C. Shih, Phys. Lett. {\bf 127B}, 242 (1983).
\bibitem{ref11} L. Van Hove, Phys. Lett. {\bf B232}, 509 (1989).
\bibitem{ref12} L. Van Hove and A. Giovannini, Z. Phys. {\bf C30}, 391 (1988).
\bibitem{ref13} Z. Koba, H.B. Nielson and P. Olessen, Nucl. Phys. {\bf B40},
   317 (1972).
\bibitem{ref14} P. Carruthers and I. Sarcevic, Phys. Lett. {\bf B189}, 
   442 (1987).
\bibitem{ref15} M. Ga\'{z}dzicki and S. Mr\'{o}wczynski, Z. Phys. {\bf C54},
   127 (1992).
\bibitem{ref16} M. Stephanov, K. Rajagopal, E. Shuryak, Phys. Rev. Lett.
   {\bf 81}, 4816 (1998).
\bibitem{ref17} M.A. Haiasz, A.D. Jackson, R.E. Shrock, M.A. Stepanov,
   and J.J.M. Verbaarschot, Phys. Rev. {\bf D58}, 96007 (1998).
\bibitem{ref18} G. Baym and H. Heiselberg, Phys. Lett. {\bf B469}, 7 (1999).
\bibitem{ref19} S. Jeon and V. Koch, Phys. Rev. Lett {\bf 83}, 5435 (1999);
   Phys. Rev. Lett. {\bf 85}, 2076 (2000).
\bibitem{ref20} M. Asakawa, U. Heinz and B. M\"{u}ller, Phys. Rev. Lett.
   {\bf 85}, 2072 (2000).
\bibitem{ref21} H. Heiselberg and A.D. Jackson, Phys. Rev. {\bf C63},
   064904 (2001).
\bibitem{frag} A.Z. Mekjian and S.J. Lee, Phys. Rev. {\bf A44}, 6294 (1991).
\bibitem{canon} S.J. Lee and A.Z. Mekjian, Phys. Rev. {\bf C45}, 1284 (1992).
\bibitem{massd} S.J. Lee and A.Z. Mekjian, Phys. Rev. {\bf C45}, 1284 (1992).
\bibitem{lgpha} S.J. Lee and A.Z. Mekjian, Phys. Rev. {\bf C56}, 2621 (1997).
\bibitem{scale} S.J. Lee and A.Z. Mekjian, Phys. Rev. {\bf C47}, 2266 (1993).
\bibitem{power} S.J. Lee and A.Z. Mekjian, Phys. Rev. {\bf C50}, 3025 (1994).
\bibitem{zarecur} K.C. Chase and A.Z. Mekjian, Phys. Rev. {\bf C 49}, 2164
   (1994); {\bf 50}, 2078 (1994).
\bibitem{hiera} S. Lupia, A. Giovannini, and R. Ugoccioni,
   Z. Phys. {\bf C59}, 427 (1993).
\bibitem{refgyu}M. Gyulassy and S. Kaufmann, Phys. Rev. Lett. {\bf 40},
   298 (1978); J. Phys. {\bf A11}, 1715 (1978).
\bibitem{prl86} A.Z. Mekjian, Phys. Rev. Lett. {\bf 86}, 220 (2001).
\bibitem{feyn} R.P. Feynman, {\it Statistical Physics} (Benjamin/Cummings,
   New York, 1972).
\bibitem{prc65} A. Mekjian, Phys. Rev. {\bf C65}, 014907 (2002).
\bibitem{refmek} A.Z. Mekjian, B.R. Schlei and D. Strottman,
   Phys. Rev. {\bf C58}, 3627 (1998).
\bibitem{becat} F. Becattini, A. Giovannini and S. Lupia, Z. Phys. {\bf C72},
   43 (1996).
\bibitem{rhwa} R. Hwa, in {\it Proc. of the Santa Fe Workshop
   on Inbtermittency in High Energy Collisions} (World Scientific,
   Singapore, 1990, F. Cooper, R. Hwa and I. Sarcevic, Edit).
\bibitem{hegyi} S. Hegyi, Phys. Lett. {\bf B309}, 443 (1993);
   {\bf B318}, 642 (1993).
\bibitem{poistrn}P. Carruthers and C.C. Shih, Int. J. Modern Phys. {\bf A2},
   1447 (1987).
\bibitem{qoptic} J. Klauder and E.C.G. Sudarshan, {\it Fundamentals of
   Quantum Optics} (Benjamin, N.Y., 1968).
\bibitem{biyaj} H. Biyajima, Prog. Theo. Physics {\bf 69}, 966 (1983).
\bibitem{advnp} S. Das Gupta, A.Z. Mekjian, B. Tsang, Adv. Nuc. Phys.
   {\bf 26}, 89 (2001).  


\end{thebibliography}
\end{document}